\documentclass[superscriptaddress, twocolumn, 10pt]{revtex4-2}
\usepackage[nearskip, margin=0pt, caption=false]{subfig}
\usepackage{hyperref}
\usepackage[labelfont=bf, justification=raggedright, singlelinecheck=false, figurename={Fig.}, tablename={Table}, labelsep=period, font=small]{caption}
\usepackage{graphicx}
\usepackage{multirow}
\usepackage{amsmath,amssymb,amsfonts}
\usepackage{amsthm}
\usepackage{mathrsfs}
\usepackage{booktabs}
\usepackage{algorithm}
\usepackage{algorithmicx}
\usepackage{algpseudocode}
\usepackage{array}
\usepackage{bm}
\usepackage{siunitx}
\usepackage{empheq}
\usepackage{threeparttable}
\usepackage{ragged2e}

\makeatletter
\def\section{\@startsection
  {section}{1}{\z@}%
  {-3.0ex \@plus -0.5ex \@minus -.2ex}%
  {3.0ex \@plus .2ex}%
  {\fontsize{10}{13}\selectfont\bfseries\centering}}
\makeatother
\makeatletter
\def\subsection{\@startsection
  {subsection}{2}{\z@}%
  {-2.0ex\@plus -0.5ex \@minus -.2ex}%
  {2.0ex \@plus .2ex}%
  {\normalfont\normalsize\bfseries\centering}}
\makeatother

\usepackage{circuitikz}
\newcommand{\myDAC}[1]{ \draw (#1)  +(16.5pt,0pt) --+(6pt,12pt) --+(-16.5pt,12pt) --+(-16.5pt,-12pt) --+(6pt,-12pt) --+(16.5pt,0pt) ; \draw[] (#1)node{\footnotesize DAC~~~~}; } % #1=name
\newcommand{\myADC}[1]{ \draw (#1)  +(-16.5pt,0pt) --+(-6pt,12pt) --+(16.5pt,12pt) --+(16.5pt,-12pt) --+(-6pt,-12pt) --+(-16.5pt,0pt) ; \draw[] (#1)node[]{\footnotesize~~ADC}; } % #1 = name

\usepackage{algorithmicx}
\usepackage{algpseudocode}
\algrenewcommand{\algorithmiccomment}[1]{\hspace{0.8cm}(\textit{#1})}
\newcounter{algorithmctr}

\newcommand{\mat}[1]{\bm{\mathrm{#1}}} % for matrices

\newcounter{NoTableEntry}
\renewcommand*{\theNoTableEntry}{NTE-\the\value{NoTableEntry}}

\begin{document}

% place page number in the center at the bottom
\makeatletter
\def\ps@plain{%
  \let\@oddhead\@empty
  \let\@evenhead\@empty
  \def\@oddfoot{\hfil\thepage\hfil}
  \let\@evenfoot\@oddfoot
}
\makeatother
\pagestyle{plain}

\title{\fontsize{14}{15}\selectfont Scalable optical neural network with nonlocally coupled coherent photonic processor}% fit in one line

%%=============================================================%%

\author{Chun Ren}
\thanks{These authors contributed equally to this work.}
\affiliation{School of Engineering, The University of Tokyo, 7-3-1 Hongo, Bunkyo-ku, Tokyo, 113-8656, Japan.}

\author{Ryota Tanomura}
\thanks{These authors contributed equally to this work.}
\affiliation{School of Engineering, The University of Tokyo, 7-3-1 Hongo, Bunkyo-ku, Tokyo, 113-8656, Japan.}

\author{Kazuki Ichinose}
\affiliation{School of Engineering, The University of Tokyo, 7-3-1 Hongo, Bunkyo-ku, Tokyo, 113-8656, Japan.}

\author{Keigo Mizukami}
\affiliation{School of Engineering, The University of Tokyo, 7-3-1 Hongo, Bunkyo-ku, Tokyo, 113-8656, Japan.}

\author{Yoshitaka Taguchi}
\affiliation{School of Engineering, The University of Tokyo, 7-3-1 Hongo, Bunkyo-ku, Tokyo, 113-8656, Japan.}
\affiliation{Currently an Independent Researcher, Japan.}

\author{Taichiro Fukui}
\affiliation{School of Engineering, The University of Tokyo, 7-3-1 Hongo, Bunkyo-ku, Tokyo, 113-8656, Japan.}

\author{Yoshiaki Nakano}
\affiliation{School of Engineering, The University of Tokyo, 7-3-1 Hongo, Bunkyo-ku, Tokyo, 113-8656, Japan.}
\affiliation{Currently with Toyota Technological Institute, 2-12-1 Hisakata, Tempaku-ku, Nagoya, 468-8511, Japan.}

\author{Takuo Tanemura}\email{takuo.tanemura@tlab.t.u-tokyo.ac.jp}
\affiliation{School of Engineering, The University of Tokyo, 7-3-1 Hongo, Bunkyo-ku, Tokyo, 113-8656, Japan.}

%%=============================================================%%

\begin{abstract}
Optical neural networks (ONNs) based on programmable photonic integrated circuits (PICs) offer a promising route toward low-latency and energy-efficient deep learning. However, conventional photonic implementations of matrix–vector multiplication (MVM) rely on locally connected architectures, such as Mach–Zehnder interferometer (MZI) meshes, whose number of active components scales quadratically with matrix size, severely limiting scalability. Here, we present a scalable ONN that overcomes this limitation by exploiting the intrinsically diffractive and nonlocal nature of coherent light inside a silicon photonic chip. Our approach employs cascaded stages of multiport directional couplers (MDCs) interleaved with compact phase-shifter arrays, enabling strong nonlocal coupling among multiple optical modes. We show that an MDC-based optical unitary converter (OUC) requires only $3N$ phase shifters to achieve uniform coverage over the $N$-dimensional complex unitary group, in stark contrast to the $O(N^2)$ scaling of conventional MZI meshes. Based on the singular value decomposition, we demonstrate that an $N\times N$ MVM can be realized using only $7N$ phase shifters, breaking the traditional $O(N^2)$ scaling barrier. We experimentally implement a 32-input silicon photonic MVM chip with a tenfold reduction in active components and validate its performance on various classification tasks. Our results establish a practical pathway toward large-scale, energy-efficient, and reconfigurable photonic neural networks.
\end{abstract}

\maketitle
\thispagestyle{plain} % also add page number in the first page

\section{Introduction}\label{sec1}

Optical neural networks (ONN) based on programmable photonic integrated circuits (PICs) have emerged as a transformative technology for significantly reducing power consumption and latency in deep learning systems \cite{shen2017deep,xiao2021large,demirkiran2023electro,xu2024large,hua2025integrated,ahmed2025universal}. By leveraging mature, foundry-scale silicon photonics, recent demonstrations of large-scale matrix–vector multiplication (MVM) have underscored the potential to substantially expand computational capabilities using compact PICs \cite{hua2025integrated,ahmed2025universal}. 
However, conventional photonic MVM architectures typically require a prohibitive number of reconfigurable components to define weights, scaling at $O(N^2)$ for $N$ input ports.
For instance, coherent optical MVM processors based on cascaded Mach–Zehnder interferometers (MZIs) \cite{shen2017deep,reck1994experimental,clements2016optimal,miller2019waves}
employ $2N^2+N$ tunable phase shifters due to the singular value decomposition (SVD) of MVM into two optical unitary converters (OUCs) with $N^2$ phase shifters and an array of $N$ variable optical attenuators (VOAs). 
Similar scaling constraints also apply to incoherent MVM circuits, which generally necessitate $N^2$ discrete switches \cite{feldmann2021parallel}, weight banks \cite{tait2017neuromorphic}, VOAs \cite{ashtiani2022chip,hua2025integrated}, or weight unit cells \cite{ahmed2025universal}.
Consequently, a naive implementation of MVM results in a rapid growth of the device footprint, optical loss, and power consumption as $N$ increases, creating a severe bottleneck for the scalability of ONN PICs.

In contrast, it has been demonstrated that space-efficient ONNs are possible by exploiting the diffractive nature of coherent light. High-accuracy image classification tasks were achieved by propagating coherent light through only few layers of diffractive phase plates \cite{lin2018all,zhou2021large} or metasurfaces \cite{zheng2022meta,luo2024meta}, which provide all-to-all connections between thousands of input neurons.
Such diffractive ONNs have also been implemented on two-dimensional photonic integrated platforms \cite{zhu2022space,fu2023photonic,wu2023lithography,onodera2025arbitrary}. For example, wave propagation inside programmable multimode waveguides was utilized to achieve ONNs with up to 49 inputs \cite{onodera2025arbitrary}, which is significantly larger than those achieved by MZI-based PICs \cite{shen2017deep,zhang2021optical,pai2023experimentally,bandyopadhyay2024single,xu2024large,wu2025scaling,ikeda2023integrated}. 
These studies imply that fully connected nonlocal coupling with complex-valued weight---naturally provided by the diffractive property of light---plays a key role in realizing scalable ONNs. 
Nevertheless, these previous demonstrations relied on bulky optics \cite{lin2018all, zhou2021large, zheng2022meta,luo2024meta}, required external optical illumination \cite{wu2023lithography,onodera2025arbitrary}, 
and/or exhibited limited reconfigurability \cite{lin2018all,fu2023photonic,zheng2022meta,zhu2022space,luo2024meta}, 
leaving a significant room to explore fully integrated, energy-efficient, and reconfigurable ONN architectures.

Here, we demonstrate a scalable and reconfigurable ONN using a compact silicon photonic MVM chip with significantly fewer phase shifters by fully exploiting the coherent diffractive nature of light. Unlike conventional coherent MVM circuits based on MZI meshes, we employ cascaded stages of multiport directional couplers (MDCs) interleaved with tunable phase shifter arrays.
Owing to the strong nonlocal coherent coupling provided by each MDC \cite{tanomura2022optical,zelaya2025integrated,ren202532},
we reveal that an MDC-based OUC (MDC-OUC) with only $3N$ phase shifters already provides sufficiently uniform sampling across the entire unitary group $\mathsf{U}(N)$, in clear contrast to an MZI-based OUC (MZI-OUC) that requires $N^2$ phase shifters to attain equivalent coverage of $\mathsf{U}(N)$. 
Based on the principle of SVD, only $7N$ phase shifters in total are sufficient to implement $N\times N$ MVM to achieve various ONN tasks, breaking the conventional $O(N^2)$ scaling barrier.
We develop a 32-input silicon photonic coherent MVM chip with tenfold reduction in the total number of phase shifters to successfully demonstrate various deep learning tasks. This work represents the largest-input reconfigurable coherent MVM PIC reported to date and highlights the scalability of our scheme to larger matrix sizes with substantially fewer active components than conventional approaches.

\vspace{-10pt}% adjust spacing
\section{Results}\label{sec2}

\subsection{Architectures of ONN and MVM chip}

\begin{figure*}
\centering
\includegraphics[width=\textwidth]{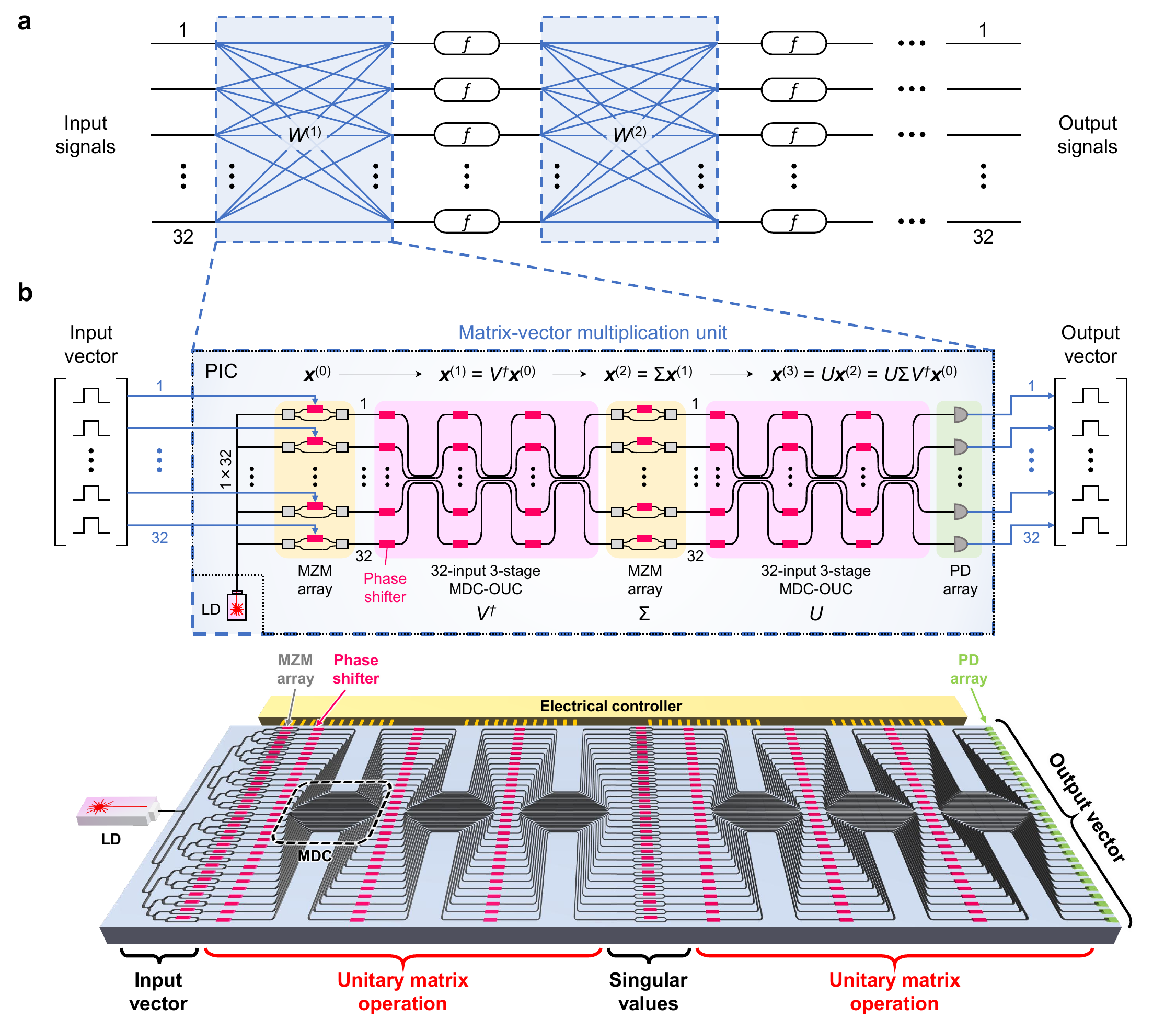}
\caption{ONN using 32-input fully integrated photonic chip.
\textbf{a} Schematic of the 32-input ONN. Each layer has a weight matrix of size $32\times32$. 
\textbf{b} Schematic of the photonic chip, where MVM is conducted entirely in the optical domain. The input vector generated by 32 MZMs is successively multiplied by a $32\times32$ unitary matrix $V^\dagger$, a diagonal singular value matrix $\Sigma$, and another $32\times32$ unitary matrix $U$, and finally detected by 32 PDs. The unitary matrices are realized by two 32-input 3-stage MDC-OUCs.
}\label{Fig1}
\end{figure*}

\begin{figure*}[p]
\centering
\includegraphics[width=\textwidth]{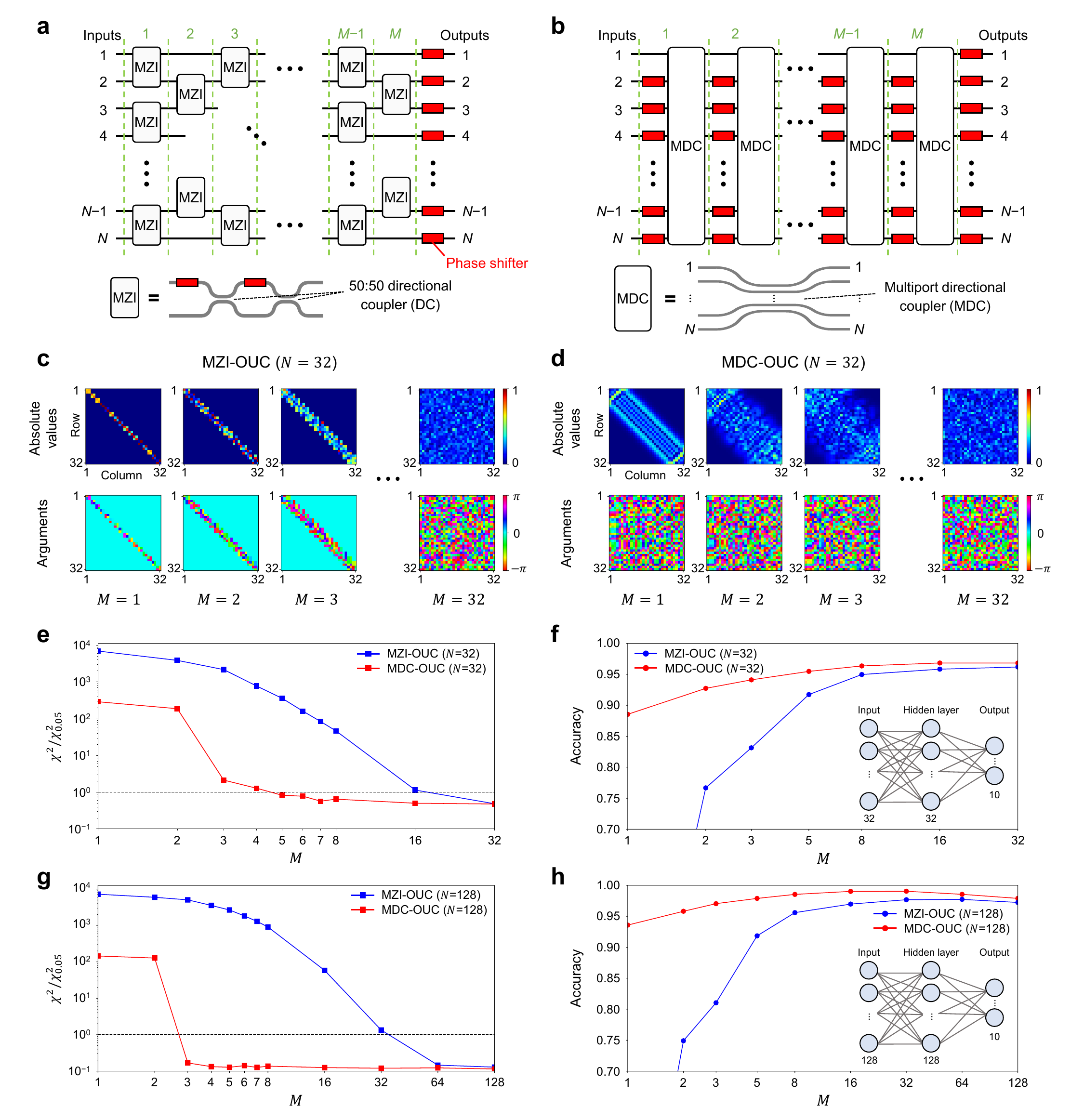}
\caption{Numerical comparison of MDC-OUC and MZI-OUC. 
\textbf{a} Schematic of $N$-input $M$-stage MZI-OUC. 
\textbf{b} Schematic of $N$-input $M$-stage MDC-OUC.
\textbf{c} $32\times32$ unitary matrices generated by MZI-OUC ($N = 32$) with various number of stages $M$. The absolute values (upper panel) and arguments (lower panel) of all components of an example matrix $U$ are plotted. 
\textbf{d} $32\times32$ unitary matrices generated by MDC-OUC ($N = 32$) with various number of stages $M$. 
\textbf{e} Haar randomness of $32\times32$ unitary matrices generated by MZI-OUC and MDC-OUC ($N = 32$) with various $M$.
\textbf{f} MNIST image classification accuracy obtained by two-layer 32-input ONN for both cases ($N = 32$).
\textbf{g} Haar randomness of $128\times128$ unitary matrices generated by MZI-OUC and MDC-OUC  ($N = 128$).
\textbf{h} MNIST image classification accuracy obtained by two-layer 128-input ONN for both cases ($N = 128$).
}\label{Fig2}
\end{figure*}

\begin{figure*}
\centering
\includegraphics[width=\textwidth]{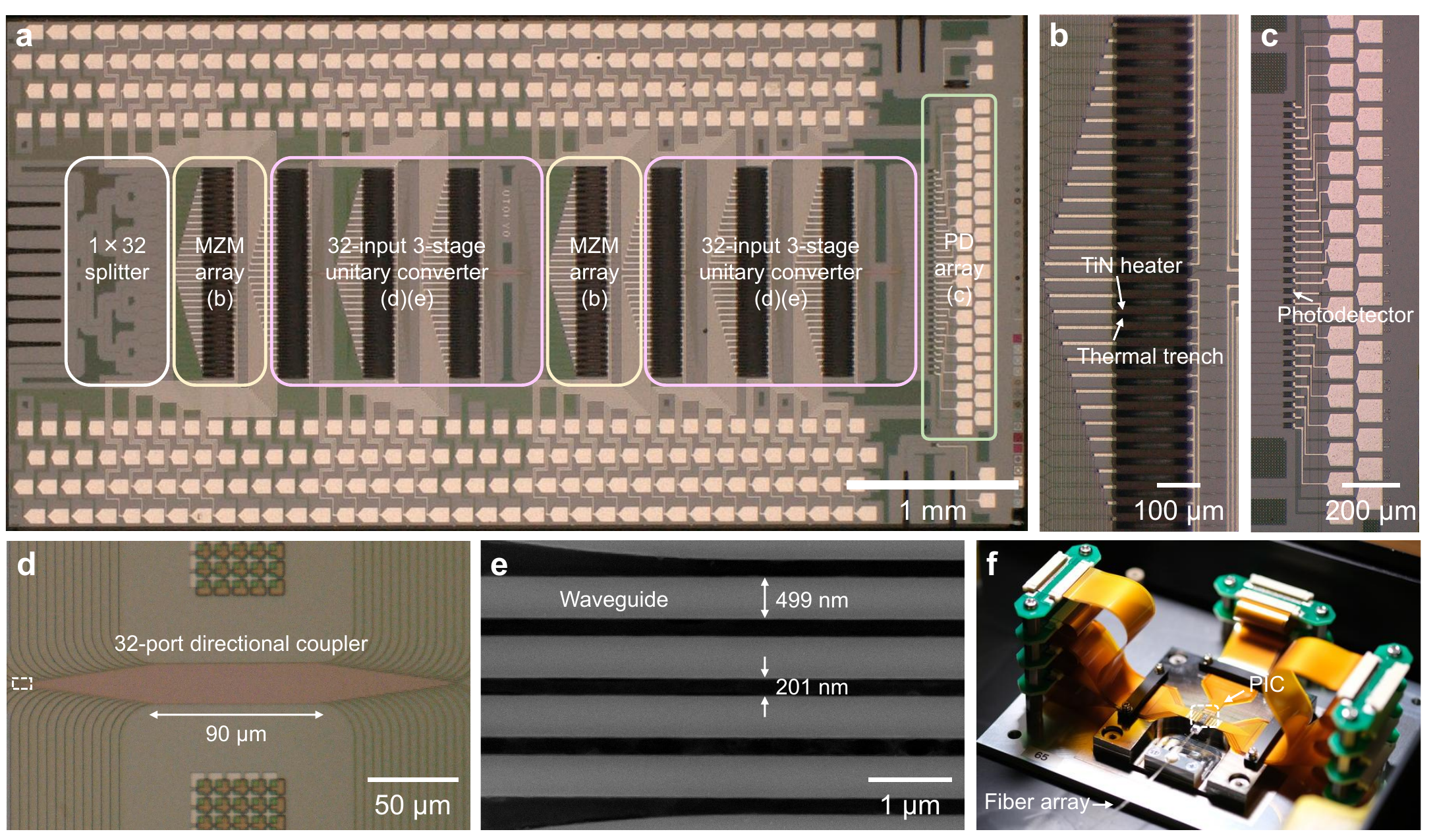}
\caption{Fabricated silicon photonic 32-input MVM PIC.
\textbf{a} Microscope image of the entire PIC. 
\textbf{b} Array of 32 MZMs. 
\textbf{c} Array of 32 PDs. 
\textbf{d} 32-port directional coupler inside an MDC-OUC.
\textbf{e} SEM image at the input of a 32-port directional coupler, marked by white broken-line box in \textbf{d}. 
\textbf{f} Packaged ONN PIC module for laboratory testing. 
}\label{Fig3}
\end{figure*}

\begin{figure*}
\centering
\includegraphics[width=\textwidth]{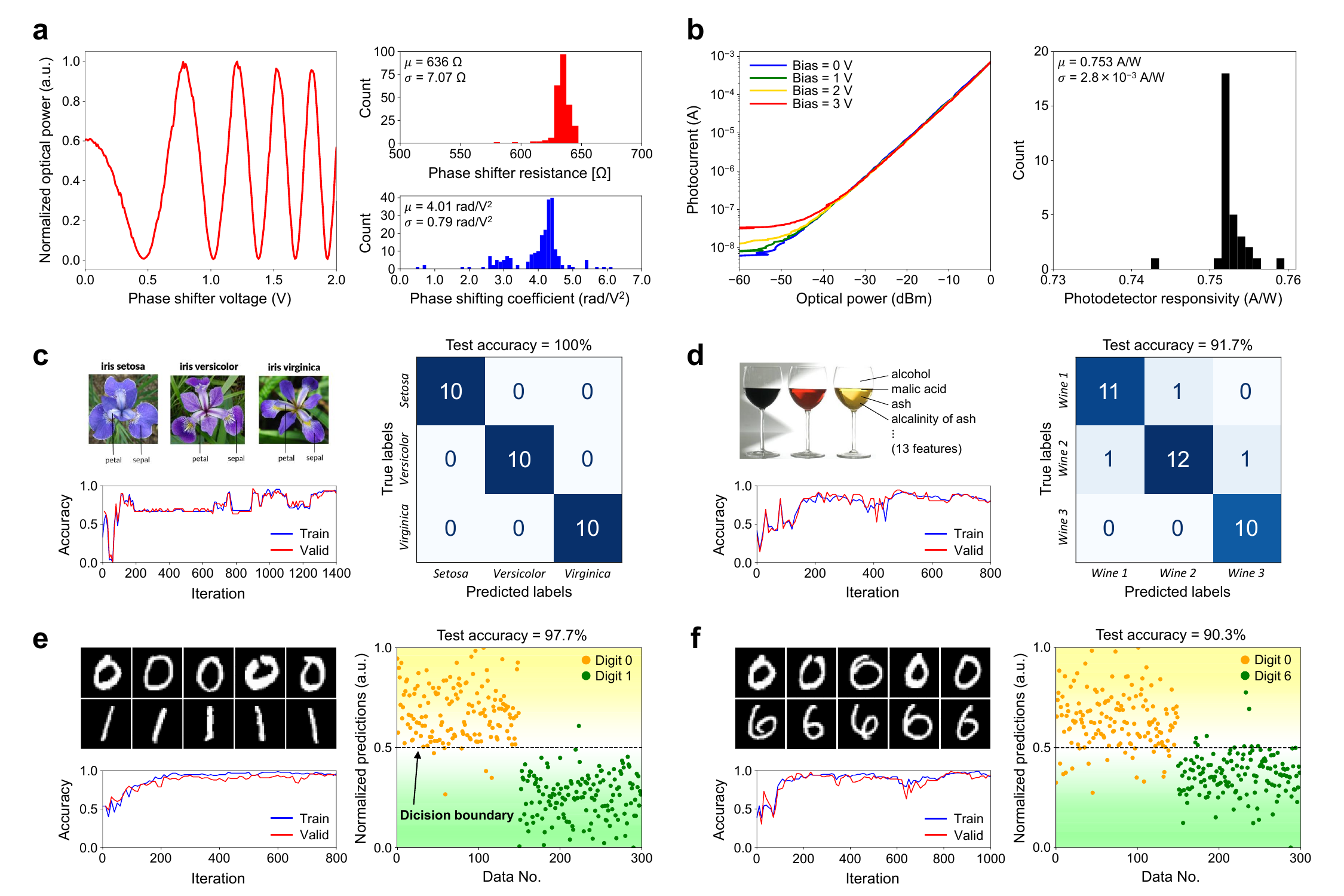}
\caption{Experimental results. 
\textbf{a} Characterization of phase shifters: output power of test MZM against driving voltage (left panel) and the histograms of the electrical resistance (right top panel) and the phase shifting coefficient (right bottom panel) measured for 226 phase shifters. In the histograms, $\mu$ and $\sigma$ denote the mean value and standard deviation, respectively. 
\textbf{b} Characterization of a test PD (left panel) and the histogram of the responsivity measured for all 32 PDs (right panel). 
\textbf{c} Experimental results of iris flower classification. Test accuracy reaches 100\% for 30 testing data out of 150 data in the whole dataset. 
\textbf{d} Experimental results of wine classification. Test accuracy reaches 91.7\% for 36 testing data out of 178 data in total. 
\textbf{e} Experimental results for binary classification of handwritten digits 0 and 1 from the MNIST dataset. A subset of input images, the training and validation accuracy curves, and the final classification output are shown. A test accuracy of 97.7\% is achieved on 300 testing samples. 
\textbf{f} Experimental results for binary classification of handwritten digits 0 and 6 from the MNIST dataset. Test accuracy of 90.3\% is achieved on 300 testing samples.
}\label{Fig4}
\end{figure*}

Figure~\ref{Fig1}a shows the architecture of the ONN using a 32-input coherent MVM chip. In each layer of the ONN, the input vector is multiplied by a $32\times32$ weight matrix $W^{(i)}$. 
The multiplication is performed in the optical domain using our proposed photonic MVM chip. 
The output vector is converted to the electrical domain, applied nonlinear activation functions $f$, and sent to the next ONN layer.

Figure~\ref{Fig1}b shows the details of the photonic MVM chip.
Coherent light from a laser diode (LD) is split uniformly to 32 channels. The input vector is encoded into optical signals by Mach-Zehnder modulators (MZMs). 
The optical signals then transmit through the $32\times32$ weight matrix multiplication unit. It is decomposed through SVD into a product of three matrices: $W^{(i)} = U \Sigma V^\dagger$, where $U$ and $V$ are unitary matrices and $\Sigma$ is a diagonal matrix consisting of singular values. Each unitary matrix is realized by an MDC-OUC, composed of a series of tunable phase shifter arrays interleaved by fixed MDCs. 
The output signals are detected by 32 photodetectors (PDs) and converted to electrical domain, where activation functions are applied through digital signal processing.

To enable full reconfigurability of MDC-OUC to implement arbitrary $32\times32$ unitary matrices, $U$ and $V^\dagger$, 32 stages of phase shifter arrays and MDCs should generally be required \cite{tang2021ten}.
This leads to $O(N^2)$ number of phase shifters, which is not scalable.
Instead, in this work, we employ a compact structure with only three stages, reducing the total number of phase shifters by a factor of $<1/10$. 
Thanks to the unique all-to-all coupling nature of MDC that offers wide coverage in the unitary space, we demonstrate both numerically and experimentally that three stages are sufficient to achieve high test accuracies in various ONN tasks.

\subsection{Contribution of nonlocal coupling to ONN performance}

The key feature of our scheme is the use of MDC-OUCs consisting of only three stages of 32-port phase shifter arrays that can uniformly sample $32\times 32$ unitary matrices covering the entire $\mathsf{U}(32)$.
This approach, therefore, allows us to maintain high ONN accuracy even with a significantly reduced number of phase shifters.

To clarify this unique advantage of MDC-OUC, we numerically compare its property with that of conventional MZI-OUC.
Figures~\ref{Fig2}a and \ref{Fig2}b illustrate the structures of MZI-OUC and MDC-OUC having $N$ input and $N$ output ports.
An MZI-OUC consists of meshes of $2\times2$ MZIs \cite{clements2016optimal}, whereas MDC-OUC employs cascaded stages of $N \times N$ directional couplers and phase shifter arrays \cite{tang2021ten}.
For both OUCs, we split the entire circuit into $M$ stages, as shown by green dotted lines, so that each stage contains $N-1$ phase shifters \cite{tanomura2022scalable}.
We then have an additional phase shifter stage at the output.
When $N=M$, the total number of phase shifters is $N^2$, corresponding to arbitrary $N\times N$ unitary converters \cite{clements2016optimal,tang2021ten}.
For the device parameters assumed in this work (see Supplementary Note \ref{sec:suppl_1A} and Fig.~\ref{Fig_S1} for details), the coupling length of the $2\times2$ directional coupler in MZI is set to \SI{18.8}{\micro m} to achieve an ideal 50:50 splitting ratio, while that of MDC is set to \SI{90.0}{\micro m} to ensure sufficient coupling among multiple ports.

Figures~\ref{Fig2}c and \ref{Fig2}d show the results of generating $32 \times 32$ matrices using MZI-OUC and MDC-OUC, respectively, with $N=32$ and different $M$ (see Supplementary Note \ref{sec:suppl_1B} and Fig.~\ref{Fig_S2} for the details).
Here, we set all phase shifts to random values and visualize the $32 \times 32$ components of the generated matrix $U$.
For the case of MZI-OUC, we can confirm that $M=32$ is required to fully populate $32\times32$ matrix.
In contrast, substantially denser matrix is generated by MDC-OUC even with small $M$. 
This difference arises from the distinct coupling natures of MDCs and MZIs; an MDC offers strong coupling among multiple waveguide modes, whereas an MZI can only provide local coupling between adjacent waveguides at each stage.

To enable quantitative comparison of MZI-OUC and MDC-OUC, Fig.~\ref{Fig2}e shows the randomness of the generated $32\times32$ matrices as a function of $M$ for both cases.
Here, the randomness of a unitary matrix is quantified based on the chi-squared statistics of probability distribution of its eigenvalues \cite{mehta2004random,mezzadri2006generate,zyczkowski1994random}. 
A smaller $\chi^2$, which describes chi-squared statistics, corresponds to a higher degree of randomness in our analysis. More specifically, when $\chi^2 < \chi_{0.05}^2$, the matrix is close to Haar random with a significance level of 5\% (see Supplementary Note \ref{sec:suppl_1C} for the details).
From Fig.~\ref{Fig2}e, we can confirm that the randomness of the unitary matrices generated by MDC-OUC rapidly improves as $M$ increases from 2 to 3, and asymptotically approaches Haar random for $M\geq3$. In contrast, for the case of MZI-OUC, $\chi^2$ reduces more slowly with $M$ and finally satisfies $\chi^2 < \chi_{0.05}^2$ at $M=N=32$.
Therefore, an MDC-OUC with a significantly reduced number of stages, e.g. $M\sim3$, can realize unitary matrices distributed over the entire unitary group $\mathsf{U}(32)$ with high uniformity.

The higher randomness of the matrices realized by MDC-OUC directly contributes to the scalability of ONN. 
We consider a two-layer 32-input ONN as shown in Fig.~\ref{Fig2}f inset and compare its performance for two cases: when the unitary conversions in the ONN are performed using MZI-OUCs and MDC-OUCs. The test accuracy of the MNIST image classification task after backpropagation training is plotted in Fig.~\ref{Fig2}f for various $M$. As expected, superior performance is obtained by MDC-OUCs at $M \le 3$. 

To compare the scalability of MDC-OUC and MZI-OUC, we plot in Figs.~\ref{Fig2}g and \ref{Fig2}h similar results obtained for $N=128$. To ensure sufficient coupling, the MDC length is increased to \SI{360}{\micro m} in these simulations.
Surprisingly, Fig.~\ref{Fig2}g reveals that a rapid drop of $\chi^2$ is present at the same threshold of $M=3$, independent of $N$ (see Supplementary Fig.~\ref{Fig_S4} for details). This result indicates that the required number of phase shifters per OUC scales as $3N$ instead of $O(N^2)$.
Consequently, markedly higher ONN performance is obtained by using MDC-OUCs than MZI-OUCs at $M=3$ as shown in Fig.~\ref{Fig2}h.

Furthermore, this high ONN accuracy at $M=3$ is maintained even at larger $N$, up to $N=512$ that we could test using our computational resource, and for more difficult tasks as well, such as Fashion MNIST (see Supplementary Note \ref{sec:suppl_1D}, Table~\ref{tab:suppl_MNIST_scalability}, and Table~\ref{tab:suppl_Fashion_MNIST_scalability} for details).
These results imply that our architecture requires only $7N$ phase shifters in total---accounting for two 3-stage MDC-OUCs and one intensity modulator array---to implement $N$-input MVM, in clear contrast to the conventional MVM PIC architectures that requires $O(N^2)$ active components.
The implication of these results is significant.
For example, 128-input MVM can be implemented using only 896 phase shifters and a total MDC length of \SI{2.16}{mm} ($=\SI{360}{\micro m} \times 6$), which can readily be integrated on a compact silicon photonic chip. 

\subsection{Experimental demonstration of 32-input ONN using a silicon photonic MVM chip}

To verify the findings in the previous section, a silicon photonic coherent 32-input MVM chip with 3-stage MDC-OUCs was fabricated.
The microscope image of the entire PIC is shown in Fig.~\ref{Fig3}a. 
Enlarged views of the 32-channel MZM and the germanium-based PD array are shown in Figs.~\ref{Fig3}b and \ref{Fig3}c, respectively. 
Each MDC-OUC was equipped with three stages of 32 phase shifters interleaved by MDCs. The MDC featured a 90-\si{\micro m}-long coupling region (Fig.~\ref{Fig3}d) with 200-nm gaps between adjacent 500-nm-wide waveguides. From the scanning electron microscopy (SEM) image (Fig.~\ref{Fig3}e), the measured gaps and widths were \SI{ 201 }{ nm } and \SI{ 499 }{ nm }, respectively, which closely matched the design specifications.
The holistic $\SI{ 5.4 }{ mm } \times \SI{ 2.8 }{ mm }$ chip with 288 active components (256 phase shifters and 32 PDs)
was packaged and connected to custom-developed electrical control circuits for experimental demonstration (Fig.~\ref{Fig3}f).
The phase shifters were characterized to show uniform properties (Fig.~\ref{Fig4}a).
Due to the deep thermal trenches introduced between the phase shifters, an efficient phase shift of $1.2~\mathrm{mW}/\pi$ was obtained. 
All 32 PDs were also characterized (Fig.~\ref{Fig4}b), showing a mean responsivity of \SI{ 0.75 }{ A/W } (see Methods and Supplementary Note \ref{sec:suppl_2}~A-C for details).

Figures~\ref{Fig4}c-f present the experimental results of testing different ONN tasks using our PIC
(see Methods and Supplementary Note \ref{sec:suppl_2D} for details of the experimental procedures).
The classification results of the iris dastaset are shown in Fig.~\ref{Fig4}c, achieving a test accuracy of 100\%.
Similarly, 91.7\% accuracy is obtained for the wine dataset (Fig.~\ref{Fig4}d).
Finally, the results of the binary classification of the digits 0/1 and 0/6 using the MNIST dataset are shown in Figs.~\ref{Fig4}e and \ref{Fig4}f, respectively. High test accuracies of 97.7\% and 90.3\% are achieved for respective cases.

\section{Discussions}\label{sec3}

\begin{table*}[ht]
\begin{threeparttable}
\caption{Comparison of reconfigurable MVM PICs.}\label{tab1:benchmark}
\begin{tabular*}{\linewidth}{@{\extracolsep{\fill}} lllllllll }
\toprule
\parbox[t]{1.0cm}{\RaggedRight Type}                      & \parbox[t]{2.2cm}{\RaggedRight Architecture}              & \parbox[t]{0.9cm}{\RaggedRight \# of inputs}      & \parbox[t]{1.6cm}{\RaggedRight \# of phase shifters} & \parbox[t]{1.5cm}{\RaggedRight Chip area [\si{mm^2}]} & \parbox[t]{2.1cm}{\RaggedRight Power consumption [W]\footnotemark[1]} & \parbox[t]{0.9cm}{\RaggedRight TOPS\footnotemark[2]} & \parbox[t]{1.0cm}{\RaggedRight TOPS/W} & \parbox[t]{0.9cm}{\RaggedRight Ref.} \\
\midrule
\multirow{10}{*}{Coherent}& \textbf{MDC-OUC}          & \textbf{32}       & \textbf{256}          &  \textbf{15.1}       & \textbf{0.24}      & \textbf{8.2}  & \textbf{35} & \parbox[t]{0.9cm}{\textbf{This work}}  \\
                          & MZI-OUC\footnotemark[3]   &  4                &  48                   &  0.74                & 0.48               & 0.13          & 0.27   & \cite{shen2017deep} \\
                          & MZI-OUC\footnotemark[3]   &  4                &  60                   &  12                  & 1.2                & 0.13          & 0.11   & \cite{pai2023experimentally} \\
                          & MZI-OUC                   &  4\footnotemark[4]&  32                   &  2.1                 & 0.64               & 0.13          & 0.20   & \cite{wu2025scaling} \\
                          & MZI-OUC\footnotemark[3]   &  6                &  56                   &  16                  & 2.9                & 0.29          & 0.098  & \cite{zhang2021optical} \\
                          & MZI-OUC                   &  6                & 132\footnotemark[5]   &  34                  & 2.0                & 0.86          & 0.43   & \cite{bandyopadhyay2024single} \\
                          & MZI-OUC                   &  8\footnotemark[6]   & 144                &  29                  & 2.6                & 0.51          & 0.20   & \cite{xu2024large} \\
                          & ODFT\footnotemark[7]      & 10                &  40                   &  6.4                 & 0.018              & 0.80          & 46     & \cite{zhu2022space} \\
                          & MZI-OUC\footnotemark[3]   & 16                & 240                   &  N/A                 & N/A                & 2.0           & N/A    & \cite{ikeda2023integrated} \\
\midrule
\multirow{3}{*}{Incoherent} & VOA (PIN)               & 30\footnotemark[8]& 66\footnotemark[9]    & 9.3                  & 3.75               & 0.55\footnotemark[10]  & 0.15   & \cite{ashtiani2022chip} \\
                            & VOA (MZI)               & 64                & 4096                  & N/A                  & 1.9                & 16            & 8.6    & \cite{hua2025integrated} \\ % lightelligence
                            & Weight unit cell        & 128               & 16384\footnotemark[11]& 349                  & 19.9               & 66            & 3.3    & \cite{ahmed2025universal} \\ % lightmatter

\botrule
\end{tabular*}
\footnotetext[1]{Total power consumed for reconfiguring weights, calculated as $7\times32=224$ phase shifters for this work, $N^2$ phase shifters for coherent OUC, $2(N^2+N)$ for coherent MVM, and $N^2$ components for incoherent MVM.}
\footnotetext[2]{Assumed $4N^2$ operations for complex-valued coherent MVM/OUC, and $2N^2$ operations ($=N^2$ MAC: multiply-and-accumulate) for real-valued incoherent MVM. For fair comparison, the clock rate of the input vectors are set to 2 GHz in all cases, assuming use of carrier-injection PIN modulators \cite{bandyopadhyay2024single,hua2025integrated}.}
\footnotetext[3]{Integrated only one OUC unit.}
\footnotetext[4]{Considered only the $4\times4$ MVM unit on the demonstrated chip.}
\footnotetext[5]{Fully integrated 3-layer ONN including both linear OUCs and nonlinear units.}
\footnotetext[6]{Considered the $8\times8$ MVM unit on the demonstrated chip.}
\footnotetext[7]{Integrated two diffractive cells to implement optical discrete Fourier transforms (ODFT) and a tunable MZI array in between them.}
\footnotetext[8]{Integrated three MVM units, each providing $30\times 4$, $4\times 3$, and $3\times 2$ MVM operation.}
\footnotetext[9]{Number of PIN VOAs.}
\footnotetext[10]{Calculated number of operations of all three MVM units.}
\footnotetext[11]{Number of weight unit cells.}
\end{threeparttable}
\end{table*}

\begin{figure}[ht]
\centering
\includegraphics[width=\columnwidth]{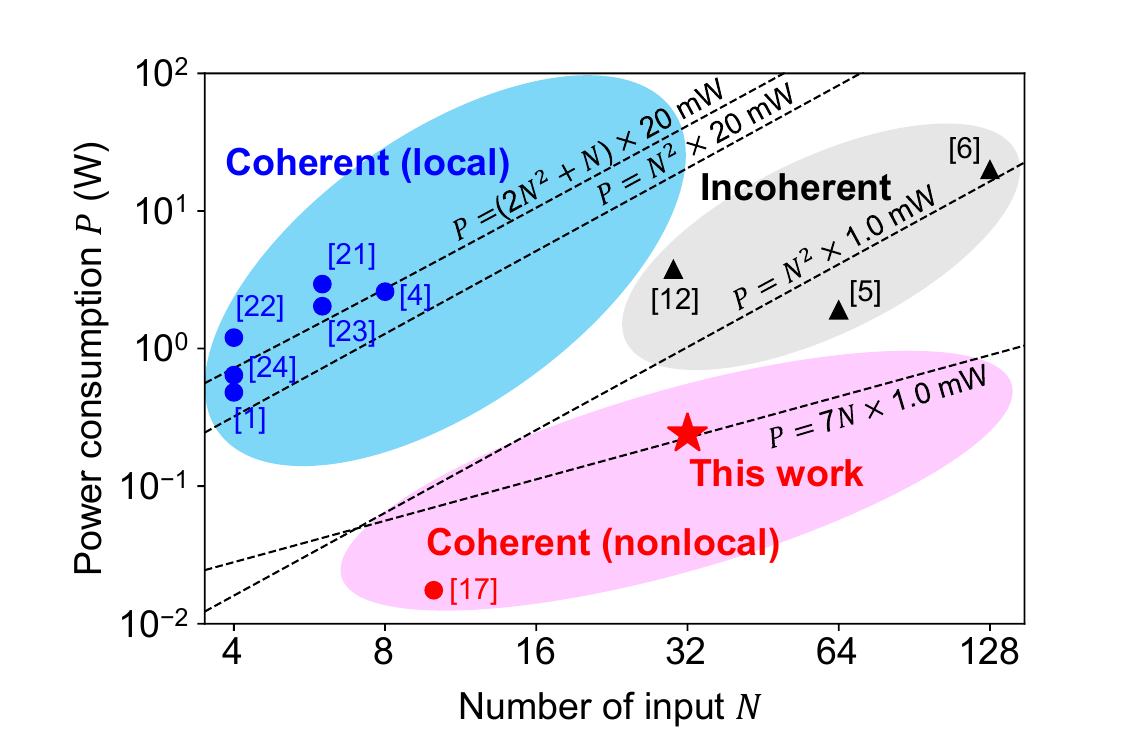}
\caption{Benchmark comparison of this work against previously reported reconfigurable MVM PICs. 
See Table~\ref{tab1:benchmark} for exact values.
}\label{Fig5}
\end{figure}

We have demonstrated scalable ONN using a fully integrated 32-input silicon photonic MVM chip based on MDC-OUCs.
Thanks to the efficient nonlocal coherent coupling provided by each MDC, the number of phase shifters was reduced by a factor of $<1/10$ without degrading the ONN accuracy.
This distinct feature was verified through comprehensive numerical analysis. 
We revealed that an MDC-OUC can quickly populate the entire $32\times32$ unitary matrix space with a nearly Haar-random distribution using only three phase-shifting stages, 
in clear contrast to a conventional MZI-OUC.

Our experimental achievement is compared with previous demonstrations of reconfigurable MVM PICs in Table~\ref{tab1:benchmark}.
By leveraging the superior scalability of MDC-OUC, this work demonstrates the largest number of input ports among coherent MVM PICs reported to date.
Each phase shifter with deep thermal trenches integrated on our chip consumes only \SI{1.05}{mW} in average, which is substantially smaller than the typical power consumption of \num{10}$\sim$\SI{20}{mW} per phase shifter required in previous devices \cite{shen2017deep,bandyopadhyay2024single,xu2024large}.
Combined with the tenfold reduction in the number of phase shifters, we achieve a total power consumption of \SI{0.27}{W} for the entire chip, of which \SI{0.24}{W} is attributed to reconfiguring the weights.
Assuming a modulation and detection rate of \SI{2}{GHz}, which is feasible by using carrier-injection-based MZMs for the input vector generator \cite{bandyopadhyay2024single,hua2025integrated}, the proposed MVM PIC can achieve a computational throughput of 8.2 tera-operation per second (TOPS). This corresponds to nearly two orders of magnitude higher energy efficiency (TOPS/W) compared to previous demonstrations using MZI-OUCs \cite{shen2017deep,wu2025scaling,zhang2021optical,pai2023experimentally,bandyopadhyay2024single,xu2024large}.

Finally, we emphasize that the unique property of our scheme remains valid at larger $N$.
From the comprehensive analysis, we have revealed that the ONN performance improves rapidly when the number of phase-shifting stages in MDC-OUC exceeds 3, which is independent of $N$.
Consequently, the total number of phase shifters in our MDC-based MVM PIC scales as $7N$, accounting for two 3-stage MDC-OUCs and one intensity modulator array. This represents a significant improvement from the $O(N^2)$ scaling of conventional architectures, including both coherent MZI meshes and incoherent approaches (Fig.~\ref{Fig5}).
Due to this unique $O(N)$ scalability, for example, 128-input MVM can be implemented using only 896 phase shifters and a total MDC length of \SI{2.16}{mm}, which can readily be integrated on a compact silicon photonic chip.
The presented approach, therefore, offers a promising route toward energy- and space-efficient large-scale ONNs using programmable coherent PICs.

\section{Methods}\label{sec:methods}

\subsection*{Chip characterization}
The MVM chip was fabricated using a multi-project wafer service  with 220-nm-thick silicon layer provided by Advanced Micro Foundry.
Thermo-optic phase shifters were implemented by 150-\SI{}{\micro m}-long TiN heaters with thermal isolation trenches. 
The characteristics of each phase shifter were measured and calibrated (Fig.~\ref{Fig4}a). The electrical resistances of 226 TiN heaters were measured to have a mean value of $\SI{ 636 }{ \Omega }$ with a standard deviation of $\SI{ 7.07 }{ \Omega } $. 
Due to unwanted short circuits and disconnections of the bonding wires, 30 out of 256 phase shifters could not be driven. Nevertheless, thanks to the all-to-all coupling feature of MDCs that provided redundancy to our chip, the effect of these non-operating phase shifters could be compensated for by tuning other phase shifters during the training process.
From the output optical power measured as a function of the applied voltage, we derived a quadratic voltage-to-phase relationship of the form $\phi = kV^2 + \text{const.}$, where $\phi$ is the phase shift, $k$ is a fitting parameter representing the phase shift coefficient, and $V$ is the applied voltage. The fitted $k$ had a mean value of \SI{ 4.01 }{ rad/V^2 } and a standard deviation of \SI{ 0.79 }{ rad/V^2 }. 
All 32 PDs were characterized by changing the input laser power (Fig.~\ref{Fig4}b). Results showed an average responsivity of \SI{0.75}{A/W} with a standard deviation of \SI{2.8e-3}{A/W}.

\subsection*{Experimental setup}
The MVM chip was mounted on a thermoelectric cooler (Thorlabs PTC1/M). 
The 256 phase shifters were driven by eight 40-channel, 16-bit digital-to-analog converters (DACs, AD5370), connected to a 256-channel voltage-follower circuit fabricated on custom printed circuit boards (see Supplementary Note \ref{sec:suppl_2B} for details). The photocurrent signals from 32 PDs were amplified by 32-channel transimpedance amplifiers (TIAs) and digitized using four 8-channel analog-to-digital converters (ADCs, MCP3564R) with 24-bit precision. Data acquisition and weight updating were done by a microcontroller unit (PIC24FJ256GB606) through serial peripheral interface (SPI) communication with the DACs/ADCs. Synchronization between setting input vectors and retrieving output signals was managed by a Python-programmed serial port module (FT232H) (see Supplementary Fig.~\ref{Fig_S11}).
Coherent light at a wavelength of \SI{1550}{nm} and optical power of 6 dBm from a tunable laser source (Keysight 81980A) was guided by a polarization-maintaining fiber and coupled to the transverse-electric (TE) mode of the silicon photonic waveguide of the MVM PIC via an edge coupler. From the characterizations of test passive waveguides, the fiber-to-chip coupling loss was estimated to be around \SI{3.2}{dB}.

\subsection*{Training algorithm}
We adopted a simulated annealing method \cite{bertsimas1993simulated} to update all phase shifters during the in-situ ONN training in the experiment. The initial values were set randomly with a uniform distribution in the range $[0, 2\pi)$. During the training process, the following optimization loop was performed to minimize the loss $L$:
\begin{enumerate}
\item Conducted forward inference of the training batch and evaluated the cross-entropy classification loss defined as $L = -\sum_{n}\sum_{i} p_{i,n} \log q_{i,n}$, where $p_{i,n}$ is the true value (equals to 1 for the correct category $i$ and 0 for the others) of the $n$-th data of the training batch, and $q_{i,n}$ is the predicted probability of the $i$-th category calculated from the inference results of the $n$-th training data.

\item A randomly selected phase shifter was perturbed by a random value between $\pm \delta\phi$, where $\delta\phi$ is an adaptive parameter updated during the training. The initial value was set as $\delta\phi = 0.1 \pi$.

\item Re-evaluated the loss $L'$ and compared it with the loss $L$ before perturbation. If $L' < L$, the perturbed phase shift was accepted. Otherwise, the perturbed phase shift was accepted only with a probability of $e^{-(L'-L)/T}$, where $T$ denotes the temperature parameter. During iterative optimization, $\delta\phi$ and $T$ were gradually decreased to ensure convergence. The loop was terminated once the condition $\delta\phi < 10^{-3}$ or $T < 10^{-10}$ was met.
\end{enumerate}

\subsection*{Dataset preparation}
We tested four different classification tasks in the experimental demonstrations. 
The complete iris dataset \cite{fisher1936fisher}, consisting of 150 instances, was divided for training (90 samples), validation (30 samples), and testing (30 samples). Each sample, containing four values, was fed to 32 input channels of our ONN with zero padding on both sides. Similarly, the wine dataset \cite{lichman2013uci} of 178 instances was divided for training (106 samples), validation (36 samples), and testing (36 samples). 
For the binary classification task using the MNIST dataset \cite{lecun2002gradient}, discrete cosine transform (DCT) \cite{ahmed1974discrete} was first applied to the original image of $28 \times 28$ pixel size. Then, the first 32 low-frequency components were extracted to reduce input dimensionality. Images corresponding to digits 0 and 1 were randomly selected to construct training, validation, and testing sets consisting of 100, 100, and 300 samples, respectively. The experiment of classifying digits 0 and 6 was conducted following the same procedure.

\section*{Data availability}
The data that support the findings in this study are included in the article and its supplementary information. Other data are available from the corresponding authors upon reasonable request.

\section*{Acknowledgements}
The authors acknowledge Go Soma, Shun Takahashi, and Rui Tang for fruitful discussions.
This work was obtained in part from Japan Society
of Promotion of Science (JSPS) KAKENHI (JP21K18168, JP23H00272, JP25KJ0805), and was supported partially by MbSC2030, the University of Tokyo.
C.R. acknowledges financial support from JSPS and World-leading Innovative Graduate Study Program - Materials Education program for the future leaders in Research, Industry, and Technology (WINGS-MERIT), the University of Tokyo.

\section*{Author contribution}
C.R., R.T., and T.T. conceived the experiments. C.R. performed comprehensive analysis, system calibration, ONN experiments, and data analysis. R.T. conceived the device concept, performed the initial numerical analysis, designed the PIC, and packaged the PIC. R.T. and K.I. characterized the basic properties of the PIC and developed the control electronics. K.M. developed the numerical models of ONN. Y.T. assisted with the experiment. T.F assisted with PIC design. Y.N. and T.T. contributed to the overall discussion and provided experimental facilities. C.R., R.T., and T.T. wrote the manuscript with inputs from all authors. T.T. supervised the project.

\section*{Competing interests}
The authors declare no competing interests.

\section*{Additional information}
\textbf{Supplementary information}
The online version contains supplementary material.

\bibliography{ref}

%apsrev4-2.bst 2019-01-14 (MD) hand-edited version of apsrev4-1.bst
%Control: key (0)
%Control: author (8) initials jnrlst
%Control: editor formatted (1) identically to author
%Control: production of article title (0) allowed
%Control: page (0) single
%Control: year (1) truncated
%Control: production of eprint (0) enabled
\begin{thebibliography}{42}%
\makeatletter
\providecommand \@ifxundefined [1]{%
 \@ifx{#1\undefined}
}%
\providecommand \@ifnum [1]{%
 \ifnum #1\expandafter \@firstoftwo
 \else \expandafter \@secondoftwo
 \fi
}%
\providecommand \@ifx [1]{%
 \ifx #1\expandafter \@firstoftwo
 \else \expandafter \@secondoftwo
 \fi
}%
\providecommand \natexlab [1]{#1}%
\providecommand \enquote  [1]{``#1''}%
\providecommand \bibnamefont  [1]{#1}%
\providecommand \bibfnamefont [1]{#1}%
\providecommand \citenamefont [1]{#1}%
\providecommand \href@noop [0]{\@secondoftwo}%
\providecommand \href [0]{\begingroup \@sanitize@url \@href}%
\providecommand \@href[1]{\@@startlink{#1}\@@href}%
\providecommand \@@href[1]{\endgroup#1\@@endlink}%
\providecommand \@sanitize@url [0]{\catcode `\\12\catcode `\$12\catcode
  `\&12\catcode `\#12\catcode `\^12\catcode `\_12\catcode `\%12\relax}%
\providecommand \@@startlink[1]{}%
\providecommand \@@endlink[0]{}%
\providecommand \url  [0]{\begingroup\@sanitize@url \@url }%
\providecommand \@url [1]{\endgroup\@href {#1}{\urlprefix }}%
\providecommand \urlprefix  [0]{URL }%
\providecommand \Eprint [0]{\href }%
\providecommand \doibase [0]{https://doi.org/}%
\providecommand \selectlanguage [0]{\@gobble}%
\providecommand \bibinfo  [0]{\@secondoftwo}%
\providecommand \bibfield  [0]{\@secondoftwo}%
\providecommand \translation [1]{[#1]}%
\providecommand \BibitemOpen [0]{}%
\providecommand \bibitemStop [0]{}%
\providecommand \bibitemNoStop [0]{.\EOS\space}%
\providecommand \EOS [0]{\spacefactor3000\relax}%
\providecommand \BibitemShut  [1]{\csname bibitem#1\endcsname}%
\let\auto@bib@innerbib\@empty
%</preamble>
\bibitem [{\citenamefont {Shen}\ \emph {et~al.}(2017)\citenamefont {Shen},
  \citenamefont {Harris}, \citenamefont {Skirlo}, \citenamefont {Prabhu},
  \citenamefont {Baehr-Jones}, \citenamefont {Hochberg}, \citenamefont {Sun},
  \citenamefont {Zhao}, \citenamefont {Larochelle}, \citenamefont {Englund}
  \emph {et~al.}}]{shen2017deep}%
  \BibitemOpen
  \bibfield  {author} {\bibinfo {author} {\bibfnamefont {Y.}~\bibnamefont
  {Shen}}, \bibinfo {author} {\bibfnamefont {N.~C.}\ \bibnamefont {Harris}},
  \bibinfo {author} {\bibfnamefont {S.}~\bibnamefont {Skirlo}}, \bibinfo
  {author} {\bibfnamefont {M.}~\bibnamefont {Prabhu}}, \bibinfo {author}
  {\bibfnamefont {T.}~\bibnamefont {Baehr-Jones}}, \bibinfo {author}
  {\bibfnamefont {M.}~\bibnamefont {Hochberg}}, \bibinfo {author}
  {\bibfnamefont {X.}~\bibnamefont {Sun}}, \bibinfo {author} {\bibfnamefont
  {S.}~\bibnamefont {Zhao}}, \bibinfo {author} {\bibfnamefont {H.}~\bibnamefont
  {Larochelle}}, \bibinfo {author} {\bibfnamefont {D.}~\bibnamefont {Englund}},
  \emph {et~al.},\ }\bibfield  {title} {\bibinfo {title} {Deep learning with
  coherent nanophotonic circuits},\ }\href@noop {} {\bibfield  {journal}
  {\bibinfo  {journal} {Nature Photonics}\ }\textbf {\bibinfo {volume} {11}},\
  \bibinfo {pages} {441} (\bibinfo {year} {2017})}\BibitemShut {NoStop}%
\bibitem [{\citenamefont {Xiao}\ \emph {et~al.}(2021)\citenamefont {Xiao},
  \citenamefont {On}, \citenamefont {Van~Vaerenbergh}, \citenamefont {Liang},
  \citenamefont {Beausoleil},\ and\ \citenamefont {Yoo}}]{xiao2021large}%
  \BibitemOpen
  \bibfield  {author} {\bibinfo {author} {\bibfnamefont {X.}~\bibnamefont
  {Xiao}}, \bibinfo {author} {\bibfnamefont {M.~B.}\ \bibnamefont {On}},
  \bibinfo {author} {\bibfnamefont {T.}~\bibnamefont {Van~Vaerenbergh}},
  \bibinfo {author} {\bibfnamefont {D.}~\bibnamefont {Liang}}, \bibinfo
  {author} {\bibfnamefont {R.~G.}\ \bibnamefont {Beausoleil}},\ and\ \bibinfo
  {author} {\bibfnamefont {S.}~\bibnamefont {Yoo}},\ }\bibfield  {title}
  {\bibinfo {title} {Large-scale and energy-efficient tensorized optical neural
  networks on {III}--{V}-on-silicon {MOSCAP} platform},\ }\href@noop {}
  {\bibfield  {journal} {\bibinfo  {journal} {APL Photonics}\ }\textbf
  {\bibinfo {volume} {6}} (\bibinfo {year} {2021})}\BibitemShut {NoStop}%
\bibitem [{\citenamefont {Demirkiran}\ \emph {et~al.}(2023)\citenamefont
  {Demirkiran}, \citenamefont {Eris}, \citenamefont {Wang}, \citenamefont
  {Elmhurst}, \citenamefont {Moore}, \citenamefont {Harris}, \citenamefont
  {Basumallik}, \citenamefont {Reddi}, \citenamefont {Joshi},\ and\
  \citenamefont {Bunandar}}]{demirkiran2023electro}%
  \BibitemOpen
  \bibfield  {author} {\bibinfo {author} {\bibfnamefont {C.}~\bibnamefont
  {Demirkiran}}, \bibinfo {author} {\bibfnamefont {F.}~\bibnamefont {Eris}},
  \bibinfo {author} {\bibfnamefont {G.}~\bibnamefont {Wang}}, \bibinfo {author}
  {\bibfnamefont {J.}~\bibnamefont {Elmhurst}}, \bibinfo {author}
  {\bibfnamefont {N.}~\bibnamefont {Moore}}, \bibinfo {author} {\bibfnamefont
  {N.~C.}\ \bibnamefont {Harris}}, \bibinfo {author} {\bibfnamefont
  {A.}~\bibnamefont {Basumallik}}, \bibinfo {author} {\bibfnamefont {V.~J.}\
  \bibnamefont {Reddi}}, \bibinfo {author} {\bibfnamefont {A.}~\bibnamefont
  {Joshi}},\ and\ \bibinfo {author} {\bibfnamefont {D.}~\bibnamefont
  {Bunandar}},\ }\bibfield  {title} {\bibinfo {title} {An electro-photonic
  system for accelerating deep neural networks},\ }\href@noop {} {\bibfield
  {journal} {\bibinfo  {journal} {ACM Journal on Emerging Technologies in
  Computing Systems}\ }\textbf {\bibinfo {volume} {19}},\ \bibinfo {pages} {1}
  (\bibinfo {year} {2023})}\BibitemShut {NoStop}%
\bibitem [{\citenamefont {Xu}\ \emph {et~al.}(2024)\citenamefont {Xu},
  \citenamefont {Zhou}, \citenamefont {Ma}, \citenamefont {Deng}, \citenamefont
  {Dai},\ and\ \citenamefont {Fang}}]{xu2024large}%
  \BibitemOpen
  \bibfield  {author} {\bibinfo {author} {\bibfnamefont {Z.}~\bibnamefont
  {Xu}}, \bibinfo {author} {\bibfnamefont {T.}~\bibnamefont {Zhou}}, \bibinfo
  {author} {\bibfnamefont {M.}~\bibnamefont {Ma}}, \bibinfo {author}
  {\bibfnamefont {C.}~\bibnamefont {Deng}}, \bibinfo {author} {\bibfnamefont
  {Q.}~\bibnamefont {Dai}},\ and\ \bibinfo {author} {\bibfnamefont
  {L.}~\bibnamefont {Fang}},\ }\bibfield  {title} {\bibinfo {title}
  {Large-scale photonic chiplet {Taichi} empowers 160-{TOPS/W} artificial
  general intelligence},\ }\href@noop {} {\bibfield  {journal} {\bibinfo
  {journal} {Science}\ }\textbf {\bibinfo {volume} {384}},\ \bibinfo {pages}
  {202} (\bibinfo {year} {2024})}\BibitemShut {NoStop}%
\bibitem [{\citenamefont {Hua}\ \emph {et~al.}(2025)\citenamefont {Hua},
  \citenamefont {Divita}, \citenamefont {Yu}, \citenamefont {Peng},
  \citenamefont {Roques-Carmes}, \citenamefont {Su}, \citenamefont {Chen},
  \citenamefont {Bai}, \citenamefont {Zou}, \citenamefont {Zhu} \emph
  {et~al.}}]{hua2025integrated}%
  \BibitemOpen
  \bibfield  {author} {\bibinfo {author} {\bibfnamefont {S.}~\bibnamefont
  {Hua}}, \bibinfo {author} {\bibfnamefont {E.}~\bibnamefont {Divita}},
  \bibinfo {author} {\bibfnamefont {S.}~\bibnamefont {Yu}}, \bibinfo {author}
  {\bibfnamefont {B.}~\bibnamefont {Peng}}, \bibinfo {author} {\bibfnamefont
  {C.}~\bibnamefont {Roques-Carmes}}, \bibinfo {author} {\bibfnamefont
  {Z.}~\bibnamefont {Su}}, \bibinfo {author} {\bibfnamefont {Z.}~\bibnamefont
  {Chen}}, \bibinfo {author} {\bibfnamefont {Y.}~\bibnamefont {Bai}}, \bibinfo
  {author} {\bibfnamefont {J.}~\bibnamefont {Zou}}, \bibinfo {author}
  {\bibfnamefont {Y.}~\bibnamefont {Zhu}}, \emph {et~al.},\ }\bibfield  {title}
  {\bibinfo {title} {An integrated large-scale photonic accelerator with
  ultralow latency},\ }\href@noop {} {\bibfield  {journal} {\bibinfo  {journal}
  {Nature}\ }\textbf {\bibinfo {volume} {640}},\ \bibinfo {pages} {361}
  (\bibinfo {year} {2025})}\BibitemShut {NoStop}%
\bibitem [{\citenamefont {Ahmed}\ \emph {et~al.}(2025)\citenamefont {Ahmed},
  \citenamefont {Baghdadi}, \citenamefont {Bernadskiy}, \citenamefont {Bowman},
  \citenamefont {Braid}, \citenamefont {Carr}, \citenamefont {Chen},
  \citenamefont {Ciccarella}, \citenamefont {Cole}, \citenamefont {Cooke} \emph
  {et~al.}}]{ahmed2025universal}%
  \BibitemOpen
  \bibfield  {author} {\bibinfo {author} {\bibfnamefont {S.~R.}\ \bibnamefont
  {Ahmed}}, \bibinfo {author} {\bibfnamefont {R.}~\bibnamefont {Baghdadi}},
  \bibinfo {author} {\bibfnamefont {M.}~\bibnamefont {Bernadskiy}}, \bibinfo
  {author} {\bibfnamefont {N.}~\bibnamefont {Bowman}}, \bibinfo {author}
  {\bibfnamefont {R.}~\bibnamefont {Braid}}, \bibinfo {author} {\bibfnamefont
  {J.}~\bibnamefont {Carr}}, \bibinfo {author} {\bibfnamefont {C.}~\bibnamefont
  {Chen}}, \bibinfo {author} {\bibfnamefont {P.}~\bibnamefont {Ciccarella}},
  \bibinfo {author} {\bibfnamefont {M.}~\bibnamefont {Cole}}, \bibinfo {author}
  {\bibfnamefont {J.}~\bibnamefont {Cooke}}, \emph {et~al.},\ }\bibfield
  {title} {\bibinfo {title} {Universal photonic artificial intelligence
  acceleration},\ }\href@noop {} {\bibfield  {journal} {\bibinfo  {journal}
  {Nature}\ }\textbf {\bibinfo {volume} {640}},\ \bibinfo {pages} {368}
  (\bibinfo {year} {2025})}\BibitemShut {NoStop}%
\bibitem [{\citenamefont {Reck}\ \emph {et~al.}(1994)\citenamefont {Reck},
  \citenamefont {Zeilinger}, \citenamefont {Bernstein},\ and\ \citenamefont
  {Bertani}}]{reck1994experimental}%
  \BibitemOpen
  \bibfield  {author} {\bibinfo {author} {\bibfnamefont {M.}~\bibnamefont
  {Reck}}, \bibinfo {author} {\bibfnamefont {A.}~\bibnamefont {Zeilinger}},
  \bibinfo {author} {\bibfnamefont {H.~J.}\ \bibnamefont {Bernstein}},\ and\
  \bibinfo {author} {\bibfnamefont {P.}~\bibnamefont {Bertani}},\ }\bibfield
  {title} {\bibinfo {title} {Experimental realization of any discrete unitary
  operator},\ }\href@noop {} {\bibfield  {journal} {\bibinfo  {journal}
  {Physical Review Letters}\ }\textbf {\bibinfo {volume} {73}},\ \bibinfo
  {pages} {58} (\bibinfo {year} {1994})}\BibitemShut {NoStop}%
\bibitem [{\citenamefont {Clements}\ \emph {et~al.}(2016)\citenamefont
  {Clements}, \citenamefont {Humphreys}, \citenamefont {Metcalf}, \citenamefont
  {Kolthammer},\ and\ \citenamefont {Walmsley}}]{clements2016optimal}%
  \BibitemOpen
  \bibfield  {author} {\bibinfo {author} {\bibfnamefont {W.~R.}\ \bibnamefont
  {Clements}}, \bibinfo {author} {\bibfnamefont {P.~C.}\ \bibnamefont
  {Humphreys}}, \bibinfo {author} {\bibfnamefont {B.~J.}\ \bibnamefont
  {Metcalf}}, \bibinfo {author} {\bibfnamefont {W.~S.}\ \bibnamefont
  {Kolthammer}},\ and\ \bibinfo {author} {\bibfnamefont {I.~A.}\ \bibnamefont
  {Walmsley}},\ }\bibfield  {title} {\bibinfo {title} {Optimal design for
  universal multiport interferometers},\ }\href@noop {} {\bibfield  {journal}
  {\bibinfo  {journal} {Optica}\ }\textbf {\bibinfo {volume} {3}},\ \bibinfo
  {pages} {1460} (\bibinfo {year} {2016})}\BibitemShut {NoStop}%
\bibitem [{\citenamefont {Miller}(2019)}]{miller2019waves}%
  \BibitemOpen
  \bibfield  {author} {\bibinfo {author} {\bibfnamefont {D.~A.}\ \bibnamefont
  {Miller}},\ }\bibfield  {title} {\bibinfo {title} {Waves, modes,
  communications, and optics: a tutorial},\ }\href@noop {} {\bibfield
  {journal} {\bibinfo  {journal} {Advances in Optics and Photonics}\ }\textbf
  {\bibinfo {volume} {11}},\ \bibinfo {pages} {679} (\bibinfo {year}
  {2019})}\BibitemShut {NoStop}%
\bibitem [{\citenamefont {Feldmann}\ \emph {et~al.}(2021)\citenamefont
  {Feldmann}, \citenamefont {Youngblood}, \citenamefont {Karpov}, \citenamefont
  {Gehring}, \citenamefont {Li}, \citenamefont {Stappers}, \citenamefont
  {Le~Gallo}, \citenamefont {Fu}, \citenamefont {Lukashchuk}, \citenamefont
  {Raja} \emph {et~al.}}]{feldmann2021parallel}%
  \BibitemOpen
  \bibfield  {author} {\bibinfo {author} {\bibfnamefont {J.}~\bibnamefont
  {Feldmann}}, \bibinfo {author} {\bibfnamefont {N.}~\bibnamefont
  {Youngblood}}, \bibinfo {author} {\bibfnamefont {M.}~\bibnamefont {Karpov}},
  \bibinfo {author} {\bibfnamefont {H.}~\bibnamefont {Gehring}}, \bibinfo
  {author} {\bibfnamefont {X.}~\bibnamefont {Li}}, \bibinfo {author}
  {\bibfnamefont {M.}~\bibnamefont {Stappers}}, \bibinfo {author}
  {\bibfnamefont {M.}~\bibnamefont {Le~Gallo}}, \bibinfo {author}
  {\bibfnamefont {X.}~\bibnamefont {Fu}}, \bibinfo {author} {\bibfnamefont
  {A.}~\bibnamefont {Lukashchuk}}, \bibinfo {author} {\bibfnamefont {A.~S.}\
  \bibnamefont {Raja}}, \emph {et~al.},\ }\bibfield  {title} {\bibinfo {title}
  {Parallel convolutional processing using an integrated photonic tensor
  core},\ }\href@noop {} {\bibfield  {journal} {\bibinfo  {journal} {Nature}\
  }\textbf {\bibinfo {volume} {589}},\ \bibinfo {pages} {52} (\bibinfo {year}
  {2021})}\BibitemShut {NoStop}%
\bibitem [{\citenamefont {Tait}\ \emph {et~al.}(2017)\citenamefont {Tait},
  \citenamefont {De~Lima}, \citenamefont {Zhou}, \citenamefont {Wu},
  \citenamefont {Nahmias}, \citenamefont {Shastri},\ and\ \citenamefont
  {Prucnal}}]{tait2017neuromorphic}%
  \BibitemOpen
  \bibfield  {author} {\bibinfo {author} {\bibfnamefont {A.~N.}\ \bibnamefont
  {Tait}}, \bibinfo {author} {\bibfnamefont {T.~F.}\ \bibnamefont {De~Lima}},
  \bibinfo {author} {\bibfnamefont {E.}~\bibnamefont {Zhou}}, \bibinfo {author}
  {\bibfnamefont {A.~X.}\ \bibnamefont {Wu}}, \bibinfo {author} {\bibfnamefont
  {M.~A.}\ \bibnamefont {Nahmias}}, \bibinfo {author} {\bibfnamefont {B.~J.}\
  \bibnamefont {Shastri}},\ and\ \bibinfo {author} {\bibfnamefont {P.~R.}\
  \bibnamefont {Prucnal}},\ }\bibfield  {title} {\bibinfo {title} {Neuromorphic
  photonic networks using silicon photonic weight banks},\ }\href@noop {}
  {\bibfield  {journal} {\bibinfo  {journal} {Scientific Reports}\ }\textbf
  {\bibinfo {volume} {7}},\ \bibinfo {pages} {7430} (\bibinfo {year}
  {2017})}\BibitemShut {NoStop}%
\bibitem [{\citenamefont {Ashtiani}\ \emph {et~al.}(2022)\citenamefont
  {Ashtiani}, \citenamefont {Geers},\ and\ \citenamefont
  {Aflatouni}}]{ashtiani2022chip}%
  \BibitemOpen
  \bibfield  {author} {\bibinfo {author} {\bibfnamefont {F.}~\bibnamefont
  {Ashtiani}}, \bibinfo {author} {\bibfnamefont {A.~J.}\ \bibnamefont
  {Geers}},\ and\ \bibinfo {author} {\bibfnamefont {F.}~\bibnamefont
  {Aflatouni}},\ }\bibfield  {title} {\bibinfo {title} {An on-chip photonic
  deep neural network for image classification},\ }\href@noop {} {\bibfield
  {journal} {\bibinfo  {journal} {Nature}\ }\textbf {\bibinfo {volume} {606}},\
  \bibinfo {pages} {501} (\bibinfo {year} {2022})}\BibitemShut {NoStop}%
\bibitem [{\citenamefont {Lin}\ \emph {et~al.}(2018)\citenamefont {Lin},
  \citenamefont {Rivenson}, \citenamefont {Yardimci}, \citenamefont {Veli},
  \citenamefont {Luo}, \citenamefont {Jarrahi},\ and\ \citenamefont
  {Ozcan}}]{lin2018all}%
  \BibitemOpen
  \bibfield  {author} {\bibinfo {author} {\bibfnamefont {X.}~\bibnamefont
  {Lin}}, \bibinfo {author} {\bibfnamefont {Y.}~\bibnamefont {Rivenson}},
  \bibinfo {author} {\bibfnamefont {N.~T.}\ \bibnamefont {Yardimci}}, \bibinfo
  {author} {\bibfnamefont {M.}~\bibnamefont {Veli}}, \bibinfo {author}
  {\bibfnamefont {Y.}~\bibnamefont {Luo}}, \bibinfo {author} {\bibfnamefont
  {M.}~\bibnamefont {Jarrahi}},\ and\ \bibinfo {author} {\bibfnamefont
  {A.}~\bibnamefont {Ozcan}},\ }\bibfield  {title} {\bibinfo {title}
  {All-optical machine learning using diffractive deep neural networks},\
  }\href@noop {} {\bibfield  {journal} {\bibinfo  {journal} {Science}\ }\textbf
  {\bibinfo {volume} {361}},\ \bibinfo {pages} {1004} (\bibinfo {year}
  {2018})}\BibitemShut {NoStop}%
\bibitem [{\citenamefont {Zhou}\ \emph {et~al.}(2021)\citenamefont {Zhou},
  \citenamefont {Lin}, \citenamefont {Wu}, \citenamefont {Chen}, \citenamefont
  {Xie}, \citenamefont {Li}, \citenamefont {Fan}, \citenamefont {Wu},
  \citenamefont {Fang},\ and\ \citenamefont {Dai}}]{zhou2021large}%
  \BibitemOpen
  \bibfield  {author} {\bibinfo {author} {\bibfnamefont {T.}~\bibnamefont
  {Zhou}}, \bibinfo {author} {\bibfnamefont {X.}~\bibnamefont {Lin}}, \bibinfo
  {author} {\bibfnamefont {J.}~\bibnamefont {Wu}}, \bibinfo {author}
  {\bibfnamefont {Y.}~\bibnamefont {Chen}}, \bibinfo {author} {\bibfnamefont
  {H.}~\bibnamefont {Xie}}, \bibinfo {author} {\bibfnamefont {Y.}~\bibnamefont
  {Li}}, \bibinfo {author} {\bibfnamefont {J.}~\bibnamefont {Fan}}, \bibinfo
  {author} {\bibfnamefont {H.}~\bibnamefont {Wu}}, \bibinfo {author}
  {\bibfnamefont {L.}~\bibnamefont {Fang}},\ and\ \bibinfo {author}
  {\bibfnamefont {Q.}~\bibnamefont {Dai}},\ }\bibfield  {title} {\bibinfo
  {title} {Large-scale neuromorphic optoelectronic computing with a
  reconfigurable diffractive processing unit},\ }\href@noop {} {\bibfield
  {journal} {\bibinfo  {journal} {Nature Photonics}\ }\textbf {\bibinfo
  {volume} {15}},\ \bibinfo {pages} {367} (\bibinfo {year} {2021})}\BibitemShut
  {NoStop}%
\bibitem [{\citenamefont {Zheng}\ \emph {et~al.}(2022)\citenamefont {Zheng},
  \citenamefont {Liu}, \citenamefont {Zhou}, \citenamefont {Kravchenko},
  \citenamefont {Huo},\ and\ \citenamefont {Valentine}}]{zheng2022meta}%
  \BibitemOpen
  \bibfield  {author} {\bibinfo {author} {\bibfnamefont {H.}~\bibnamefont
  {Zheng}}, \bibinfo {author} {\bibfnamefont {Q.}~\bibnamefont {Liu}}, \bibinfo
  {author} {\bibfnamefont {Y.}~\bibnamefont {Zhou}}, \bibinfo {author}
  {\bibfnamefont {I.~I.}\ \bibnamefont {Kravchenko}}, \bibinfo {author}
  {\bibfnamefont {Y.}~\bibnamefont {Huo}},\ and\ \bibinfo {author}
  {\bibfnamefont {J.}~\bibnamefont {Valentine}},\ }\bibfield  {title} {\bibinfo
  {title} {Meta-optic accelerators for object classifiers},\ }\href@noop {}
  {\bibfield  {journal} {\bibinfo  {journal} {Science Advances}\ }\textbf
  {\bibinfo {volume} {8}},\ \bibinfo {pages} {eabo6410} (\bibinfo {year}
  {2022})}\BibitemShut {NoStop}%
\bibitem [{\citenamefont {Luo}\ \emph {et~al.}(2024)\citenamefont {Luo},
  \citenamefont {Xu}, \citenamefont {Xiao}, \citenamefont {Tsang},
  \citenamefont {Shu},\ and\ \citenamefont {Huang}}]{luo2024meta}%
  \BibitemOpen
  \bibfield  {author} {\bibinfo {author} {\bibfnamefont {M.}~\bibnamefont
  {Luo}}, \bibinfo {author} {\bibfnamefont {T.}~\bibnamefont {Xu}}, \bibinfo
  {author} {\bibfnamefont {S.}~\bibnamefont {Xiao}}, \bibinfo {author}
  {\bibfnamefont {H.~K.}\ \bibnamefont {Tsang}}, \bibinfo {author}
  {\bibfnamefont {C.}~\bibnamefont {Shu}},\ and\ \bibinfo {author}
  {\bibfnamefont {C.}~\bibnamefont {Huang}},\ }\bibfield  {title} {\bibinfo
  {title} {Meta-optics based parallel convolutional processing for neural
  network accelerator},\ }\href@noop {} {\bibfield  {journal} {\bibinfo
  {journal} {Laser \& Photonics Reviews}\ }\textbf {\bibinfo {volume} {18}},\
  \bibinfo {pages} {2300984} (\bibinfo {year} {2024})}\BibitemShut {NoStop}%
\bibitem [{\citenamefont {Zhu}\ \emph {et~al.}(2022)\citenamefont {Zhu},
  \citenamefont {Zou}, \citenamefont {Zhang}, \citenamefont {Shi},
  \citenamefont {Luo}, \citenamefont {Wang}, \citenamefont {Cai}, \citenamefont
  {Wan}, \citenamefont {Wang}, \citenamefont {Jiang} \emph
  {et~al.}}]{zhu2022space}%
  \BibitemOpen
  \bibfield  {author} {\bibinfo {author} {\bibfnamefont {H.}~\bibnamefont
  {Zhu}}, \bibinfo {author} {\bibfnamefont {J.}~\bibnamefont {Zou}}, \bibinfo
  {author} {\bibfnamefont {H.}~\bibnamefont {Zhang}}, \bibinfo {author}
  {\bibfnamefont {Y.}~\bibnamefont {Shi}}, \bibinfo {author} {\bibfnamefont
  {S.}~\bibnamefont {Luo}}, \bibinfo {author} {\bibfnamefont {N.}~\bibnamefont
  {Wang}}, \bibinfo {author} {\bibfnamefont {H.}~\bibnamefont {Cai}}, \bibinfo
  {author} {\bibfnamefont {L.}~\bibnamefont {Wan}}, \bibinfo {author}
  {\bibfnamefont {B.}~\bibnamefont {Wang}}, \bibinfo {author} {\bibfnamefont
  {X.}~\bibnamefont {Jiang}}, \emph {et~al.},\ }\bibfield  {title} {\bibinfo
  {title} {Space-efficient optical computing with an integrated chip
  diffractive neural network},\ }\href@noop {} {\bibfield  {journal} {\bibinfo
  {journal} {Nature Communications}\ }\textbf {\bibinfo {volume} {13}},\
  \bibinfo {pages} {1044} (\bibinfo {year} {2022})}\BibitemShut {NoStop}%
\bibitem [{\citenamefont {Fu}\ \emph {et~al.}(2023)\citenamefont {Fu},
  \citenamefont {Zang}, \citenamefont {Huang}, \citenamefont {Du},
  \citenamefont {Huang}, \citenamefont {Hu}, \citenamefont {Chen},
  \citenamefont {Yang},\ and\ \citenamefont {Chen}}]{fu2023photonic}%
  \BibitemOpen
  \bibfield  {author} {\bibinfo {author} {\bibfnamefont {T.}~\bibnamefont
  {Fu}}, \bibinfo {author} {\bibfnamefont {Y.}~\bibnamefont {Zang}}, \bibinfo
  {author} {\bibfnamefont {Y.}~\bibnamefont {Huang}}, \bibinfo {author}
  {\bibfnamefont {Z.}~\bibnamefont {Du}}, \bibinfo {author} {\bibfnamefont
  {H.}~\bibnamefont {Huang}}, \bibinfo {author} {\bibfnamefont
  {C.}~\bibnamefont {Hu}}, \bibinfo {author} {\bibfnamefont {M.}~\bibnamefont
  {Chen}}, \bibinfo {author} {\bibfnamefont {S.}~\bibnamefont {Yang}},\ and\
  \bibinfo {author} {\bibfnamefont {H.}~\bibnamefont {Chen}},\ }\bibfield
  {title} {\bibinfo {title} {Photonic machine learning with on-chip diffractive
  optics},\ }\href@noop {} {\bibfield  {journal} {\bibinfo  {journal} {Nature
  Communications}\ }\textbf {\bibinfo {volume} {14}},\ \bibinfo {pages} {70}
  (\bibinfo {year} {2023})}\BibitemShut {NoStop}%
\bibitem [{\citenamefont {Wu}\ \emph {et~al.}(2023)\citenamefont {Wu},
  \citenamefont {Menarini}, \citenamefont {Gao},\ and\ \citenamefont
  {Feng}}]{wu2023lithography}%
  \BibitemOpen
  \bibfield  {author} {\bibinfo {author} {\bibfnamefont {T.}~\bibnamefont
  {Wu}}, \bibinfo {author} {\bibfnamefont {M.}~\bibnamefont {Menarini}},
  \bibinfo {author} {\bibfnamefont {Z.}~\bibnamefont {Gao}},\ and\ \bibinfo
  {author} {\bibfnamefont {L.}~\bibnamefont {Feng}},\ }\bibfield  {title}
  {\bibinfo {title} {Lithography-free reconfigurable integrated photonic
  processor},\ }\href@noop {} {\bibfield  {journal} {\bibinfo  {journal}
  {Nature Photonics}\ }\textbf {\bibinfo {volume} {17}},\ \bibinfo {pages}
  {710} (\bibinfo {year} {2023})}\BibitemShut {NoStop}%
\bibitem [{\citenamefont {Onodera}\ \emph {et~al.}(2025)\citenamefont
  {Onodera}, \citenamefont {Stein}, \citenamefont {Ash}, \citenamefont
  {Sohoni}, \citenamefont {Bosch}, \citenamefont {Yanagimoto}, \citenamefont
  {Jankowski}, \citenamefont {McKenna}, \citenamefont {Wang}, \citenamefont
  {Shvets} \emph {et~al.}}]{onodera2025arbitrary}%
  \BibitemOpen
  \bibfield  {author} {\bibinfo {author} {\bibfnamefont {T.}~\bibnamefont
  {Onodera}}, \bibinfo {author} {\bibfnamefont {M.~M.}\ \bibnamefont {Stein}},
  \bibinfo {author} {\bibfnamefont {B.~A.}\ \bibnamefont {Ash}}, \bibinfo
  {author} {\bibfnamefont {M.~M.}\ \bibnamefont {Sohoni}}, \bibinfo {author}
  {\bibfnamefont {M.}~\bibnamefont {Bosch}}, \bibinfo {author} {\bibfnamefont
  {R.}~\bibnamefont {Yanagimoto}}, \bibinfo {author} {\bibfnamefont
  {M.}~\bibnamefont {Jankowski}}, \bibinfo {author} {\bibfnamefont {T.~P.}\
  \bibnamefont {McKenna}}, \bibinfo {author} {\bibfnamefont {T.}~\bibnamefont
  {Wang}}, \bibinfo {author} {\bibfnamefont {G.}~\bibnamefont {Shvets}}, \emph
  {et~al.},\ }\bibfield  {title} {\bibinfo {title} {Arbitrary control over
  multimode wave propagation for machine learning},\ }\href@noop {} {\bibfield
  {journal} {\bibinfo  {journal} {Nature Physics}\ ,\ \bibinfo {pages} {1}}
  (\bibinfo {year} {2025})}\BibitemShut {NoStop}%
\bibitem [{\citenamefont {Zhang}\ \emph {et~al.}(2021)\citenamefont {Zhang},
  \citenamefont {Gu}, \citenamefont {Jiang}, \citenamefont {Thompson},
  \citenamefont {Cai}, \citenamefont {Paesani}, \citenamefont {Santagati},
  \citenamefont {Laing}, \citenamefont {Zhang}, \citenamefont {Yung} \emph
  {et~al.}}]{zhang2021optical}%
  \BibitemOpen
  \bibfield  {author} {\bibinfo {author} {\bibfnamefont {H.}~\bibnamefont
  {Zhang}}, \bibinfo {author} {\bibfnamefont {M.}~\bibnamefont {Gu}}, \bibinfo
  {author} {\bibfnamefont {X.}~\bibnamefont {Jiang}}, \bibinfo {author}
  {\bibfnamefont {J.}~\bibnamefont {Thompson}}, \bibinfo {author}
  {\bibfnamefont {H.}~\bibnamefont {Cai}}, \bibinfo {author} {\bibfnamefont
  {S.}~\bibnamefont {Paesani}}, \bibinfo {author} {\bibfnamefont
  {R.}~\bibnamefont {Santagati}}, \bibinfo {author} {\bibfnamefont
  {A.}~\bibnamefont {Laing}}, \bibinfo {author} {\bibfnamefont
  {Y.}~\bibnamefont {Zhang}}, \bibinfo {author} {\bibfnamefont {M.-H.}\
  \bibnamefont {Yung}}, \emph {et~al.},\ }\bibfield  {title} {\bibinfo {title}
  {An optical neural chip for implementing complex-valued neural network},\
  }\href@noop {} {\bibfield  {journal} {\bibinfo  {journal} {Nature
  Communications}\ }\textbf {\bibinfo {volume} {12}},\ \bibinfo {pages} {457}
  (\bibinfo {year} {2021})}\BibitemShut {NoStop}%
\bibitem [{\citenamefont {Pai}\ \emph {et~al.}(2023)\citenamefont {Pai},
  \citenamefont {Sun}, \citenamefont {Hughes}, \citenamefont {Park},
  \citenamefont {Bartlett}, \citenamefont {Williamson}, \citenamefont {Minkov},
  \citenamefont {Milanizadeh}, \citenamefont {Abebe}, \citenamefont
  {Morichetti} \emph {et~al.}}]{pai2023experimentally}%
  \BibitemOpen
  \bibfield  {author} {\bibinfo {author} {\bibfnamefont {S.}~\bibnamefont
  {Pai}}, \bibinfo {author} {\bibfnamefont {Z.}~\bibnamefont {Sun}}, \bibinfo
  {author} {\bibfnamefont {T.~W.}\ \bibnamefont {Hughes}}, \bibinfo {author}
  {\bibfnamefont {T.}~\bibnamefont {Park}}, \bibinfo {author} {\bibfnamefont
  {B.}~\bibnamefont {Bartlett}}, \bibinfo {author} {\bibfnamefont {I.~A.}\
  \bibnamefont {Williamson}}, \bibinfo {author} {\bibfnamefont
  {M.}~\bibnamefont {Minkov}}, \bibinfo {author} {\bibfnamefont
  {M.}~\bibnamefont {Milanizadeh}}, \bibinfo {author} {\bibfnamefont
  {N.}~\bibnamefont {Abebe}}, \bibinfo {author} {\bibfnamefont
  {F.}~\bibnamefont {Morichetti}}, \emph {et~al.},\ }\bibfield  {title}
  {\bibinfo {title} {Experimentally realized in situ backpropagation for deep
  learning in photonic neural networks},\ }\href@noop {} {\bibfield  {journal}
  {\bibinfo  {journal} {Science}\ }\textbf {\bibinfo {volume} {380}},\ \bibinfo
  {pages} {398} (\bibinfo {year} {2023})}\BibitemShut {NoStop}%
\bibitem [{\citenamefont {Bandyopadhyay}\ \emph {et~al.}(2024)\citenamefont
  {Bandyopadhyay}, \citenamefont {Sludds}, \citenamefont {Krastanov},
  \citenamefont {Hamerly}, \citenamefont {Harris}, \citenamefont {Bunandar},
  \citenamefont {Streshinsky}, \citenamefont {Hochberg},\ and\ \citenamefont
  {Englund}}]{bandyopadhyay2024single}%
  \BibitemOpen
  \bibfield  {author} {\bibinfo {author} {\bibfnamefont {S.}~\bibnamefont
  {Bandyopadhyay}}, \bibinfo {author} {\bibfnamefont {A.}~\bibnamefont
  {Sludds}}, \bibinfo {author} {\bibfnamefont {S.}~\bibnamefont {Krastanov}},
  \bibinfo {author} {\bibfnamefont {R.}~\bibnamefont {Hamerly}}, \bibinfo
  {author} {\bibfnamefont {N.}~\bibnamefont {Harris}}, \bibinfo {author}
  {\bibfnamefont {D.}~\bibnamefont {Bunandar}}, \bibinfo {author}
  {\bibfnamefont {M.}~\bibnamefont {Streshinsky}}, \bibinfo {author}
  {\bibfnamefont {M.}~\bibnamefont {Hochberg}},\ and\ \bibinfo {author}
  {\bibfnamefont {D.}~\bibnamefont {Englund}},\ }\bibfield  {title} {\bibinfo
  {title} {Single-chip photonic deep neural network with forward-only
  training},\ }\href@noop {} {\bibfield  {journal} {\bibinfo  {journal} {Nature
  Photonics}\ }\textbf {\bibinfo {volume} {18}},\ \bibinfo {pages} {1335}
  (\bibinfo {year} {2024})}\BibitemShut {NoStop}%
\bibitem [{\citenamefont {Wu}\ \emph {et~al.}(2025)\citenamefont {Wu},
  \citenamefont {Huang}, \citenamefont {Zhang}, \citenamefont {Zhou},
  \citenamefont {Wang}, \citenamefont {Dong},\ and\ \citenamefont
  {Zhang}}]{wu2025scaling}%
  \BibitemOpen
  \bibfield  {author} {\bibinfo {author} {\bibfnamefont {B.}~\bibnamefont
  {Wu}}, \bibinfo {author} {\bibfnamefont {C.}~\bibnamefont {Huang}}, \bibinfo
  {author} {\bibfnamefont {J.}~\bibnamefont {Zhang}}, \bibinfo {author}
  {\bibfnamefont {H.}~\bibnamefont {Zhou}}, \bibinfo {author} {\bibfnamefont
  {Y.}~\bibnamefont {Wang}}, \bibinfo {author} {\bibfnamefont {J.}~\bibnamefont
  {Dong}},\ and\ \bibinfo {author} {\bibfnamefont {X.}~\bibnamefont {Zhang}},\
  }\bibfield  {title} {\bibinfo {title} {Scaling up for end-to-end on-chip
  photonic neural network inference},\ }\href@noop {} {\bibfield  {journal}
  {\bibinfo  {journal} {Light: Science \& Applications}\ }\textbf {\bibinfo
  {volume} {14}},\ \bibinfo {pages} {328} (\bibinfo {year} {2025})}\BibitemShut
  {NoStop}%
\bibitem [{\citenamefont {Ikeda}\ \emph {et~al.}(2023)\citenamefont {Ikeda},
  \citenamefont {Kita}, \citenamefont {Nozaki}, \citenamefont {Takata},
  \citenamefont {Aoyama}, \citenamefont {Suzuki}, \citenamefont {Maegami},
  \citenamefont {Ohno}, \citenamefont {Cong}, \citenamefont {Yamamoto} \emph
  {et~al.}}]{ikeda2023integrated}%
  \BibitemOpen
  \bibfield  {author} {\bibinfo {author} {\bibfnamefont {K.}~\bibnamefont
  {Ikeda}}, \bibinfo {author} {\bibfnamefont {S.}~\bibnamefont {Kita}},
  \bibinfo {author} {\bibfnamefont {K.}~\bibnamefont {Nozaki}}, \bibinfo
  {author} {\bibfnamefont {K.}~\bibnamefont {Takata}}, \bibinfo {author}
  {\bibfnamefont {K.}~\bibnamefont {Aoyama}}, \bibinfo {author} {\bibfnamefont
  {K.}~\bibnamefont {Suzuki}}, \bibinfo {author} {\bibfnamefont
  {Y.}~\bibnamefont {Maegami}}, \bibinfo {author} {\bibfnamefont
  {M.}~\bibnamefont {Ohno}}, \bibinfo {author} {\bibfnamefont {G.}~\bibnamefont
  {Cong}}, \bibinfo {author} {\bibfnamefont {N.}~\bibnamefont {Yamamoto}},
  \emph {et~al.},\ }\bibfield  {title} {\bibinfo {title} {Integrated
  16$\times$16 photonic analog vector-matrix multiplier with task-specific
  tuning after deterministic calibration},\ }in\ \href@noop {} {\emph {\bibinfo
  {booktitle} {2023 49th European Conference on Optical Communications (ECOC
  2023)}}},\ Vol.\ \bibinfo {volume} {2023}\ (\bibinfo {organization} {IET},\
  \bibinfo {year} {2023})\ pp.\ \bibinfo {pages} {88--91}\BibitemShut {NoStop}%
\bibitem [{\citenamefont {Tanomura}\ \emph
  {et~al.}(2022{\natexlab{a}})\citenamefont {Tanomura}, \citenamefont
  {Mizukami}, \citenamefont {Tang}, \citenamefont {Soma}, \citenamefont
  {Tanemura},\ and\ \citenamefont {Nakano}}]{tanomura2022optical}%
  \BibitemOpen
  \bibfield  {author} {\bibinfo {author} {\bibfnamefont {R.}~\bibnamefont
  {Tanomura}}, \bibinfo {author} {\bibfnamefont {K.}~\bibnamefont {Mizukami}},
  \bibinfo {author} {\bibfnamefont {R.}~\bibnamefont {Tang}}, \bibinfo {author}
  {\bibfnamefont {G.}~\bibnamefont {Soma}}, \bibinfo {author} {\bibfnamefont
  {T.}~\bibnamefont {Tanemura}},\ and\ \bibinfo {author} {\bibfnamefont
  {Y.}~\bibnamefont {Nakano}},\ }\bibfield  {title} {\bibinfo {title} {Optical
  neural network with reduced phase shifters using multi-plane light
  conversion},\ }in\ \href@noop {} {\emph {\bibinfo {booktitle} {2022 27th
  OptoElectronics and Communications Conference (OECC) and 2022 International
  Conference on Photonics in Switching and Computing (PSC)}}}\ (\bibinfo
  {organization} {IEEE},\ \bibinfo {year} {2022})\ pp.\ \bibinfo {pages}
  {1--3}\BibitemShut {NoStop}%
\bibitem [{\citenamefont {Zelaya}\ \emph {et~al.}(2025)\citenamefont {Zelaya},
  \citenamefont {Honari-Latifpour},\ and\ \citenamefont
  {Miri}}]{zelaya2025integrated}%
  \BibitemOpen
  \bibfield  {author} {\bibinfo {author} {\bibfnamefont {K.}~\bibnamefont
  {Zelaya}}, \bibinfo {author} {\bibfnamefont {M.}~\bibnamefont
  {Honari-Latifpour}},\ and\ \bibinfo {author} {\bibfnamefont {M.-A.}\
  \bibnamefont {Miri}},\ }\bibfield  {title} {\bibinfo {title} {Integrated
  photonic programmable random matrix generator with minimal active
  components},\ }\href@noop {} {\bibfield  {journal} {\bibinfo  {journal} {npj
  Nanophotonics}\ }\textbf {\bibinfo {volume} {2}},\ \bibinfo {pages} {6}
  (\bibinfo {year} {2025})}\BibitemShut {NoStop}%
\bibitem [{\citenamefont {Ren}\ \emph {et~al.}(2025)\citenamefont {Ren},
  \citenamefont {Tanomura}, \citenamefont {Ichinose}, \citenamefont {Nakano},\
  and\ \citenamefont {Tanemura}}]{ren202532}%
  \BibitemOpen
  \bibfield  {author} {\bibinfo {author} {\bibfnamefont {C.}~\bibnamefont
  {Ren}}, \bibinfo {author} {\bibfnamefont {R.}~\bibnamefont {Tanomura}},
  \bibinfo {author} {\bibfnamefont {K.}~\bibnamefont {Ichinose}}, \bibinfo
  {author} {\bibfnamefont {Y.}~\bibnamefont {Nakano}},\ and\ \bibinfo {author}
  {\bibfnamefont {T.}~\bibnamefont {Tanemura}},\ }\bibfield  {title} {\bibinfo
  {title} {32-input optical neural network chip based on multi-plane light
  conversion},\ }in\ \href@noop {} {\emph {\bibinfo {booktitle} {2025
  Conference on Lasers and Electro-Optics (CLEO)}}}\ (\bibinfo {organization}
  {IEEE},\ \bibinfo {year} {2025})\ pp.\ \bibinfo {pages} {1--2}\BibitemShut
  {NoStop}%
\bibitem [{\citenamefont {Tang}\ \emph {et~al.}(2021)\citenamefont {Tang},
  \citenamefont {Tanomura}, \citenamefont {Tanemura},\ and\ \citenamefont
  {Nakano}}]{tang2021ten}%
  \BibitemOpen
  \bibfield  {author} {\bibinfo {author} {\bibfnamefont {R.}~\bibnamefont
  {Tang}}, \bibinfo {author} {\bibfnamefont {R.}~\bibnamefont {Tanomura}},
  \bibinfo {author} {\bibfnamefont {T.}~\bibnamefont {Tanemura}},\ and\
  \bibinfo {author} {\bibfnamefont {Y.}~\bibnamefont {Nakano}},\ }\bibfield
  {title} {\bibinfo {title} {Ten-port unitary optical processor on a silicon
  photonic chip},\ }\href@noop {} {\bibfield  {journal} {\bibinfo  {journal}
  {ACS Photonics}\ }\textbf {\bibinfo {volume} {8}},\ \bibinfo {pages} {2074}
  (\bibinfo {year} {2021})}\BibitemShut {NoStop}%
\bibitem [{\citenamefont {Tanomura}\ \emph
  {et~al.}(2022{\natexlab{b}})\citenamefont {Tanomura}, \citenamefont {Tang},
  \citenamefont {Umezaki}, \citenamefont {Soma}, \citenamefont {Tanemura},\
  and\ \citenamefont {Nakano}}]{tanomura2022scalable}%
  \BibitemOpen
  \bibfield  {author} {\bibinfo {author} {\bibfnamefont {R.}~\bibnamefont
  {Tanomura}}, \bibinfo {author} {\bibfnamefont {R.}~\bibnamefont {Tang}},
  \bibinfo {author} {\bibfnamefont {T.}~\bibnamefont {Umezaki}}, \bibinfo
  {author} {\bibfnamefont {G.}~\bibnamefont {Soma}}, \bibinfo {author}
  {\bibfnamefont {T.}~\bibnamefont {Tanemura}},\ and\ \bibinfo {author}
  {\bibfnamefont {Y.}~\bibnamefont {Nakano}},\ }\bibfield  {title} {\bibinfo
  {title} {Scalable and robust photonic integrated unitary converter based on
  multiplane light conversion},\ }\href@noop {} {\bibfield  {journal} {\bibinfo
   {journal} {Physical Review Applied}\ }\textbf {\bibinfo {volume} {17}},\
  \bibinfo {pages} {024071} (\bibinfo {year} {2022}{\natexlab{b}})}\BibitemShut
  {NoStop}%
\bibitem [{\citenamefont {Mehta}(2004)}]{mehta2004random}%
  \BibitemOpen
  \bibfield  {author} {\bibinfo {author} {\bibfnamefont {M.~L.}\ \bibnamefont
  {Mehta}},\ }\href@noop {} {\emph {\bibinfo {title} {Random matrices}}},\
  Vol.\ \bibinfo {volume} {142}\ (\bibinfo  {publisher} {Elsevier},\ \bibinfo
  {year} {2004})\BibitemShut {NoStop}%
\bibitem [{\citenamefont {Mezzadri}(2006)}]{mezzadri2006generate}%
  \BibitemOpen
  \bibfield  {author} {\bibinfo {author} {\bibfnamefont {F.}~\bibnamefont
  {Mezzadri}},\ }\bibfield  {title} {\bibinfo {title} {How to generate random
  matrices from the classical compact groups},\ }\href@noop {} {\bibfield
  {journal} {\bibinfo  {journal} {arXiv preprint math-ph/0609050}\ } (\bibinfo
  {year} {2006})}\BibitemShut {NoStop}%
\bibitem [{\citenamefont {Zyczkowski}\ and\ \citenamefont
  {Kus}(1994)}]{zyczkowski1994random}%
  \BibitemOpen
  \bibfield  {author} {\bibinfo {author} {\bibfnamefont {K.}~\bibnamefont
  {Zyczkowski}}\ and\ \bibinfo {author} {\bibfnamefont {M.}~\bibnamefont
  {Kus}},\ }\bibfield  {title} {\bibinfo {title} {Random unitary matrices},\
  }\href@noop {} {\bibfield  {journal} {\bibinfo  {journal} {Journal of Physics
  A: Mathematical and General}\ }\textbf {\bibinfo {volume} {27}},\ \bibinfo
  {pages} {4235} (\bibinfo {year} {1994})}\BibitemShut {NoStop}%
\bibitem [{\citenamefont {Bertsimas}\ and\ \citenamefont
  {Tsitsiklis}(1993)}]{bertsimas1993simulated}%
  \BibitemOpen
  \bibfield  {author} {\bibinfo {author} {\bibfnamefont {D.}~\bibnamefont
  {Bertsimas}}\ and\ \bibinfo {author} {\bibfnamefont {J.}~\bibnamefont
  {Tsitsiklis}},\ }\bibfield  {title} {\bibinfo {title} {Simulated annealing},\
  }\href@noop {} {\bibfield  {journal} {\bibinfo  {journal} {Statistical
  Science}\ }\textbf {\bibinfo {volume} {8}},\ \bibinfo {pages} {10} (\bibinfo
  {year} {1993})}\BibitemShut {NoStop}%
\bibitem [{\citenamefont {Fisher}\ and\ \citenamefont
  {Marshall}(1936)}]{fisher1936fisher}%
  \BibitemOpen
  \bibfield  {author} {\bibinfo {author} {\bibfnamefont {R.~A.}\ \bibnamefont
  {Fisher}}\ and\ \bibinfo {author} {\bibfnamefont {M.}~\bibnamefont
  {Marshall}},\ }\bibfield  {title} {\bibinfo {title} {Iris data set},\
  }\href@noop {} {\bibfield  {journal} {\bibinfo  {journal} {{UCI} Machine
  Learning Repository}\ }\textbf {\bibinfo {volume} {440}} (\bibinfo {year}
  {1936})}\BibitemShut {NoStop}%
\bibitem [{\citenamefont {Lichman}\ \emph {et~al.}(2013)\citenamefont {Lichman}
  \emph {et~al.}}]{lichman2013uci}%
  \BibitemOpen
  \bibfield  {author} {\bibinfo {author} {\bibfnamefont {M.}~\bibnamefont
  {Lichman}} \emph {et~al.},\ }\href@noop {} {\bibinfo {title} {{UCI} machine
  learning repository}},\ \bibinfo {howpublished}
  {\url{https://archive.ics.uci.edu/ml}} (\bibinfo {year} {2013})\BibitemShut
  {NoStop}%
\bibitem [{\citenamefont {LeCun}\ \emph {et~al.}(2002)\citenamefont {LeCun},
  \citenamefont {Bottou}, \citenamefont {Bengio},\ and\ \citenamefont
  {Haffner}}]{lecun2002gradient}%
  \BibitemOpen
  \bibfield  {author} {\bibinfo {author} {\bibfnamefont {Y.}~\bibnamefont
  {LeCun}}, \bibinfo {author} {\bibfnamefont {L.}~\bibnamefont {Bottou}},
  \bibinfo {author} {\bibfnamefont {Y.}~\bibnamefont {Bengio}},\ and\ \bibinfo
  {author} {\bibfnamefont {P.}~\bibnamefont {Haffner}},\ }\bibfield  {title}
  {\bibinfo {title} {Gradient-based learning applied to document recognition},\
  }\href@noop {} {\bibfield  {journal} {\bibinfo  {journal} {Proceedings of the
  IEEE}\ }\textbf {\bibinfo {volume} {86}},\ \bibinfo {pages} {2278} (\bibinfo
  {year} {2002})}\BibitemShut {NoStop}%
\bibitem [{\citenamefont {Ahmed}\ \emph {et~al.}(1974)\citenamefont {Ahmed},
  \citenamefont {Natarajan},\ and\ \citenamefont {Rao}}]{ahmed1974discrete}%
  \BibitemOpen
  \bibfield  {author} {\bibinfo {author} {\bibfnamefont {N.}~\bibnamefont
  {Ahmed}}, \bibinfo {author} {\bibfnamefont {T.}~\bibnamefont {Natarajan}},\
  and\ \bibinfo {author} {\bibfnamefont {K.~R.}\ \bibnamefont {Rao}},\
  }\bibfield  {title} {\bibinfo {title} {Discrete cosine transform},\
  }\href@noop {} {\bibfield  {journal} {\bibinfo  {journal} {IEEE Transactions
  on Computers}\ }\textbf {\bibinfo {volume} {100}},\ \bibinfo {pages} {90}
  (\bibinfo {year} {1974})}\BibitemShut {NoStop}%
\bibitem [{\citenamefont {Haus}\ \emph {et~al.}(1987)\citenamefont {Haus},
  \citenamefont {Huang}, \citenamefont {Kawakakmi},\ and\ \citenamefont
  {Whitaker}}]{haus1987coupled}%
  \BibitemOpen
  \bibfield  {author} {\bibinfo {author} {\bibfnamefont {H.~A.}\ \bibnamefont
  {Haus}}, \bibinfo {author} {\bibfnamefont {W.~P.}\ \bibnamefont {Huang}},
  \bibinfo {author} {\bibfnamefont {S.}~\bibnamefont {Kawakakmi}},\ and\
  \bibinfo {author} {\bibfnamefont {N.~A.}\ \bibnamefont {Whitaker}},\
  }\bibfield  {title} {\bibinfo {title} {Coupled-mode theory of optical
  waveguides},\ }\href@noop {} {\bibfield  {journal} {\bibinfo  {journal}
  {Journal of Lightwave Technology}\ }\textbf {\bibinfo {volume} {5}},\
  \bibinfo {pages} {16} (\bibinfo {year} {1987})}\BibitemShut {NoStop}%
\bibitem [{\citenamefont {Russell}\ \emph {et~al.}(2017)\citenamefont
  {Russell}, \citenamefont {Chakhmakhchyan}, \citenamefont {O’Brien},\ and\
  \citenamefont {Laing}}]{russell2017direct}%
  \BibitemOpen
  \bibfield  {author} {\bibinfo {author} {\bibfnamefont {N.~J.}\ \bibnamefont
  {Russell}}, \bibinfo {author} {\bibfnamefont {L.}~\bibnamefont
  {Chakhmakhchyan}}, \bibinfo {author} {\bibfnamefont {J.~L.}\ \bibnamefont
  {O’Brien}},\ and\ \bibinfo {author} {\bibfnamefont {A.}~\bibnamefont
  {Laing}},\ }\bibfield  {title} {\bibinfo {title} {Direct dialling of haar
  random unitary matrices},\ }\href@noop {} {\bibfield  {journal} {\bibinfo
  {journal} {New Journal of Physics}\ }\textbf {\bibinfo {volume} {19}},\
  \bibinfo {pages} {033007} (\bibinfo {year} {2017})}\BibitemShut {NoStop}%
\bibitem [{\citenamefont {Pai}\ \emph {et~al.}(2019)\citenamefont {Pai},
  \citenamefont {Bartlett}, \citenamefont {Solgaard},\ and\ \citenamefont
  {Miller}}]{pai2019matrix}%
  \BibitemOpen
  \bibfield  {author} {\bibinfo {author} {\bibfnamefont {S.}~\bibnamefont
  {Pai}}, \bibinfo {author} {\bibfnamefont {B.}~\bibnamefont {Bartlett}},
  \bibinfo {author} {\bibfnamefont {O.}~\bibnamefont {Solgaard}},\ and\
  \bibinfo {author} {\bibfnamefont {D.~A.}\ \bibnamefont {Miller}},\ }\bibfield
   {title} {\bibinfo {title} {Matrix optimization on universal unitary photonic
  devices},\ }\href@noop {} {\bibfield  {journal} {\bibinfo  {journal}
  {Physical Review Applied}\ }\textbf {\bibinfo {volume} {11}},\ \bibinfo
  {pages} {064044} (\bibinfo {year} {2019})}\BibitemShut {NoStop}%
\bibitem [{\citenamefont {Xiao}\ \emph {et~al.}(2017)\citenamefont {Xiao},
  \citenamefont {Rasul},\ and\ \citenamefont {Vollgraf}}]{xiao2017fashion}%
  \BibitemOpen
  \bibfield  {author} {\bibinfo {author} {\bibfnamefont {H.}~\bibnamefont
  {Xiao}}, \bibinfo {author} {\bibfnamefont {K.}~\bibnamefont {Rasul}},\ and\
  \bibinfo {author} {\bibfnamefont {R.}~\bibnamefont {Vollgraf}},\ }\bibfield
  {title} {\bibinfo {title} {Fashion-mnist: a novel image dataset for
  benchmarking machine learning algorithms},\ }\href@noop {} {\bibfield
  {journal} {\bibinfo  {journal} {arXiv preprint arXiv:1708.07747}\ } (\bibinfo
  {year} {2017})}\BibitemShut {NoStop}%
\end{thebibliography}%

%==================================================================%
%                         Supplementary                            %
%==================================================================%
\clearpage
\onecolumngrid
\appendix

% Supplementary heading
\begin{center}
{\large {\rm \it Supplementary Information for\\}
\vspace*{0.1em}
\Large \bf Scalable optical neural network with nonlocally coupled coherent \\photonic processor}
\end{center}

% Authors
\begin{center}
{\normalsize
Chun Ren,$^{1,*}$ 
Ryota Tanomura,$^{1}$ 
Kazuki Ichinose,$^{1}$ 
Keigo Mizukami,$^{1}$ 
Yoshitaka Taguchi,$^{1,2}$ \\
Taichiro Fukui,$^{1}$ 
Yoshiaki Nakano,$^{1,3}$ 
and
Takuo Tanemura$^{1,\dagger}$ 
}
\\
\vspace{1.0em}
{\small \textit{
$^{1}$School of Engineering, The University of Tokyo, Tokyo, 113-8656, Japan.\\
$^{2}$Currently an Independent Researcher, Japan.\\
$^{3}$Currently with Toyota Technological Institute, Nagoya, 468-8511, Japan. \\
}
\vspace{0.5em}
{\small
* \href{mailto:chun.ren@tlab.t.u-tokyo.ac.jp}{chun.ren@tlab.t.u-tokyo.ac.jp}\par
$\dagger$ \href{mailto:takuo.tanemura@tlab.t.u-tokyo.ac.jp}{takuo.tanemura@tlab.t.u-tokyo.ac.jp}
}
}
\end{center}
\vspace{0.3em}

% narrow margins between sections and subsections
\makeatletter
% -------- section -------- %
\renewcommand{\section}[1]{%
  \refstepcounter{section}%
  \setcounter{subsection}{0}%
  \vspace{7.0ex}% space above section
  \begin{center}%
  \bfseries\small
  \thesection.\ \MakeUppercase{#1}% all capital section header
  \end{center}%
  \vspace{2.0ex}% space below section
}
% -------- subsection -------- %
\renewcommand{\subsection}[1]{%
  \refstepcounter{subsection}%
  \vspace{5.0ex} % space above subsection
  \begin{center}%
  \bfseries\normalsize
  \thesubsection.\ #1%
  \end{center}%
  \vspace{1.5ex}% space below subsection
}
\makeatother

\renewcommand{\appendixname}{Supplementary Note}

\setcounter{equation}{0}
\setcounter{figure}{0}
\setcounter{table}{0}
\setcounter{section}{0}

\renewcommand{\theequation}{S\arabic{equation}}
\renewcommand{\thefigure}{S\arabic{figure}}
\renewcommand{\thetable}{S\arabic{table}}
\renewcommand{\thesection}{\arabic{section}}
\renewcommand{\thesubsection}{\Alph{subsection}}

\section{Numerical analysis of OUC randomness and ONN scalability}
\label{sec:suppl_1}

The performance of MDC-OUC and MZI-OUC and their randomnesses were analyzed by the following procedures.

\subsection{Transfer matrix of a multiport directional coupler}
\label{sec:suppl_1A}

Optical coupling between two waveguides is expressed by the coupled-mode equation (CME) \cite{haus1987coupled} as
\begin{empheq}[left={\empheqlbrace~}]{alignat=2}
	\begin{split}
		& c_{11} \frac{ \partial a_1 }{ \partial z } + c_{12} \frac{ \partial a_2 }{ \partial z } \exp{ \left[ j( \beta_2 -\beta_1 )z \right] } - j \kappa_{11} a_1 - j \kappa_{12} a_2 \exp{ \left[ j( \beta_2 -\beta_1 )z \right] } = 0
        \\
        & c_{22} \frac{ \partial a_2 }{ \partial z } + c_{21} \frac{ \partial a_1 }{ \partial z } \exp{ \left[ -j( \beta_2 - \beta_1 )z \right] } - j \kappa_{22} a_2 - j \kappa_{21} a_1 \exp{ \left[ - j( \beta_2 - \beta_1 )z \right] } = 0
	\end{split}
    ~,
	\label{eq:suppl_coupled_mode_eq}
\end{empheq}
where $a_1(z)$ and $a_2(z)$ are the complex amplitudes of the modes propagating inside the two waveguides. $\beta_1$ and $\beta_2$ are their propagation coefficients. The coefficients $c_{ij}$ and $\kappa_{ij}$ are defined as
\begin{align}
	c_{ ij }
		&\equiv \int_{-\infty}^{+\infty} \int_{-\infty}^{+\infty}
				\bm{z} \cdot ( \bm{E}_j \times \bm{H}_i^* + \bm{E}_i^* \times \bm{H}_j )
			~ dx dy
    ~,
	\\
	\kappa_{ ij }
		&\equiv \int_{-\infty}^{+\infty} \int_{-\infty}^{+\infty}
				\omega \varepsilon_0 ( n^2 - n_j^2 ) \bm{E}_i^* \cdot \bm{E}_j
			~ dx dy
    ~.
\end{align}
Here, $n(x,y)$ is the lateral refractive index distribution of both waveguides, whereas $n_j(x,y)$ denotes that of only the $j$-th waveguide. $\bm{z}$ is the unit vector along the $z$ axis, $\omega$ is the optical angular frequency, and $\varepsilon_0$ is the permittivity in vacuum. Here we assume that all waveguides are uniform, i.e. $\beta_1 = \beta_2$, so that Eq.~(\ref{eq:suppl_coupled_mode_eq}) is reduced to a simple matrix representation. 

In a general case of $N$-port MDC, the CME for all modes can be written as
\begin{equation}
	\mat{C} \frac{ \partial \bm{a} }{ \partial z }
	= j \mat{K} \bm{a} ~,
    \label{eq:CME}
\end{equation}
where $\bm{a} \equiv \begin{pmatrix} a_1, a_2, \dots, a_N \end{pmatrix}$,
$ \mat{C} \equiv \left( c_{ij} \right) $
and $ \mat{K} \equiv \left( \kappa_{ij} \right) $.
Through diagonalization $ \mat{C}^{-1} \mat{K} = \mat{V} \mat{\Lambda} \mat{V}^{-1} $, Eq.~(\ref{eq:CME}) can be solved as
\begin{equation}
    \frac{ \partial \left( \mat{V}^{-1} \bm{a} \right) }{ \partial z }
    =
    j \mat{\Lambda} \left( \mat{V}^{-1} \bm{a} \right)
    ~~~\Rightarrow~~~
    \bm{a}(z)
    = \mat{V} \exp{\left( j \mat{\Lambda} z \right)} \mat{V}^{-1} \bm{a}(0)
    \equiv \mat{M} \bm{a}(0)
    ~,
\end{equation}
where the diagonal matrix $\mat{\Lambda}$ contains propagation coefficients $\lambda_i$ of all eigenmodes of MDC:
\begin{equation}
	\mat{ \Lambda }
	=
	\begin{pmatrix}
		\lambda_1  &  0  &  \cdots  &  0
		\\
		0  & \lambda_2  &  \cdots  &  0
		\\
		\vdots &  \vdots  &  \ddots  &  \vdots
		\\
		0  &  0  &  \cdots  &  \lambda_N
	\end{pmatrix}
	\hspace{1pt} , \hspace{10pt}
	\exp{ \left( j \mat{\Lambda} z \right) }
	=
	\begin{pmatrix}
		e^{ j \lambda_1 z }  &  0  &  \cdots  &  0
		\\
		0  & e^{ j \lambda_2 z }  &  \cdots  &  0
		\\
		\vdots &  \vdots  &  \ddots  &  \vdots
		\\
		0  &  0  &  \cdots  &  e^{ j \lambda_N z }
	\end{pmatrix}
    ~.
\end{equation}
$\mat{M}$ represents the transfer matrix of MDC  and expressed as
\begin{equation}
    \mat{M} \equiv \mat{V} \exp{\left( j \mat{\Lambda} z \right)} \mat{V}^{-1}
    \label{eq:suppl_MDC_transfer_matrix}
    .
\end{equation}

\begin{figure*}
\centering
\includegraphics[width=\textwidth]{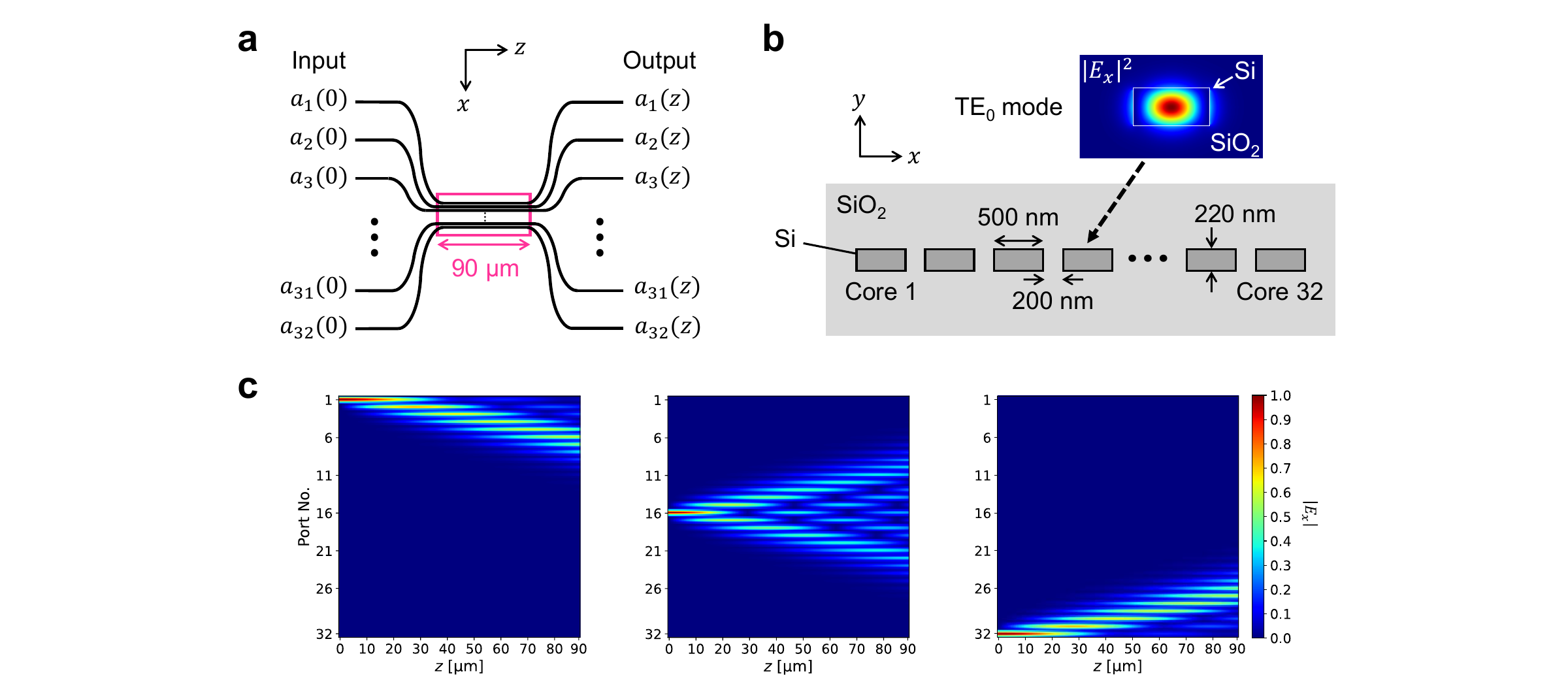}
\caption{
\textbf{a} Top view of a 32-port MDC. 
\textbf{b} Cross-sectional waveguide structure of 32-port MDC. The electric field of the $\mathrm{TE}_{0}$ mode of a waveguide is also shown.
\textbf{c} Simulated optical wave propagation at 1550-nm wavelength inside the MDC from different input ports.
}\label{Fig_S1}
\end{figure*}

Based on this formalization, we simulated the propagation inside an MDC. The structure is shown in Figs.~\ref{Fig_S1}a and \ref{Fig_S1}b. The electric field of the fundamental transverse-electric ($\mathrm{TE}_0$) mode, which is required to construct matrices $\mat{C}$ and $\mat{K}$, is derived by the finite difference eigenmode (FDE) method using Ansys Lumerical. The simulated results of optical wave propagation at 1550-nm wavelength from the 1st, 16th, and 32nd input ports are shown in Fig.~\ref{Fig_S1}c. We also conducted eigenmode expansion (EME) simulation using Ansys Lumerical and confirmed excellent agreement with these results.

\subsection{Optical unitary converters}
\label{sec:suppl_1B}

As shown in Figs.~\ref{Fig2}a-d in the main text, properties of MZI-OUCs and MDC-OUCs with $N$ input ports and $M$ phase-shifting stages are compared by analyzing random unitary matrices generated in each configuration.
The transfer matrix of an MZI operating on the $p$-th and $q$-th port is expressed using two phase parameters $\theta$ and $\phi$ as
\begin{equation}
	T_{ p, q }
	=
	j e^{ j \frac{ \theta }{ 2 } }
	\begin{pmatrix}
		1  &  \cdots  &  0  &  \cdots  &  0  &  \cdots  &  0
		\\
		\vdots  & \ddots &  \vdots   &   &  \vdots  &  &  \vdots
		\\
		0 &  \cdots &  e^{ j \phi } \sin{ \frac{ \theta }{ 2 } }  & \cdots & e^{ j \phi } \cos{ \frac{ \theta }{ 2 } } & \cdots &  0
		\\
		\vdots &   &  \vdots &  \ddots  &  \vdots &   &  \vdots
		\\
		0 &  \cdots &  \cos{ \frac{ \theta }{ 2 } }   &  \cdots  &  - \sin{ \frac{ \theta }{ 2 } }  &  \cdots  &  0
		\\
		\vdots &   &  \vdots &   &  \vdots &  \ddots   &  \vdots
		\\
		0  &  \cdots  &  0  &  \cdots  &  0  &  \cdots  &  1
	\end{pmatrix}
	~,
\end{equation}
where $(p,p)$, $(p,q)$, $(q,p)$, and $(q,q)$ elements represent the unitary coupling between two ports. 
The transfer matrix of a phase shifter array at the output is expressed as
$ \mat{\Phi}_{M+1} \equiv \mathrm{diag}\begin{pmatrix} e^{j\phi_1}, \dots, e^{j\phi_N} \end{pmatrix} $, where $\phi_1, \dots, \phi_N$ are the phase shifts at respective ports. 
Thus, the transfer matrix of the entire MZI-OUC is given by
\begin{equation}
    \mat{U}_\mathrm{MZI} = \mat{\Phi}_{M+1} \prod_{p,q} T_{p,q} ~.
    \label{eq:U_MZI}
\end{equation}

On the other hand, the transfer matrix of an MDC-OUC is described using $\mat{M}$ derived in Eq.~(\ref{eq:suppl_MDC_transfer_matrix}) as
\begin{equation}
    \mat{U}_\mathrm{MDC} = \mat{\Phi}_{M+1} \cdot \mat{M} \cdot \mat{\Phi}_{M} \cdot \mat{M} \cdot \mat{\Phi}_{M-1} \cdots \mat{M} \cdot \Phi_1 ~.
    \label{eq:U_MDC}
\end{equation}

Using Eq.~(\ref{eq:U_MZI}) and Eq.~(\ref{eq:U_MDC}), random unitary matrices are calculated for each configuration.
In the case of MZI-OUC, we set phase shifter values at each MZI following the probability functions provided in \cite{russell2017direct,pai2019matrix} to ensure that Haar randomness is obtained when $M = N$. For MDC-OUC, we set all phase shifters to random values in the range of $[0,2\pi)$ with uniform probability.
The MDC length is set to \SI{ 90 }{ \micro m }, which is consistent with Fig.~\ref{Fig_S1} and that used in our fabricated device.

\begin{figure*}
\centering
\includegraphics[width=\textwidth]{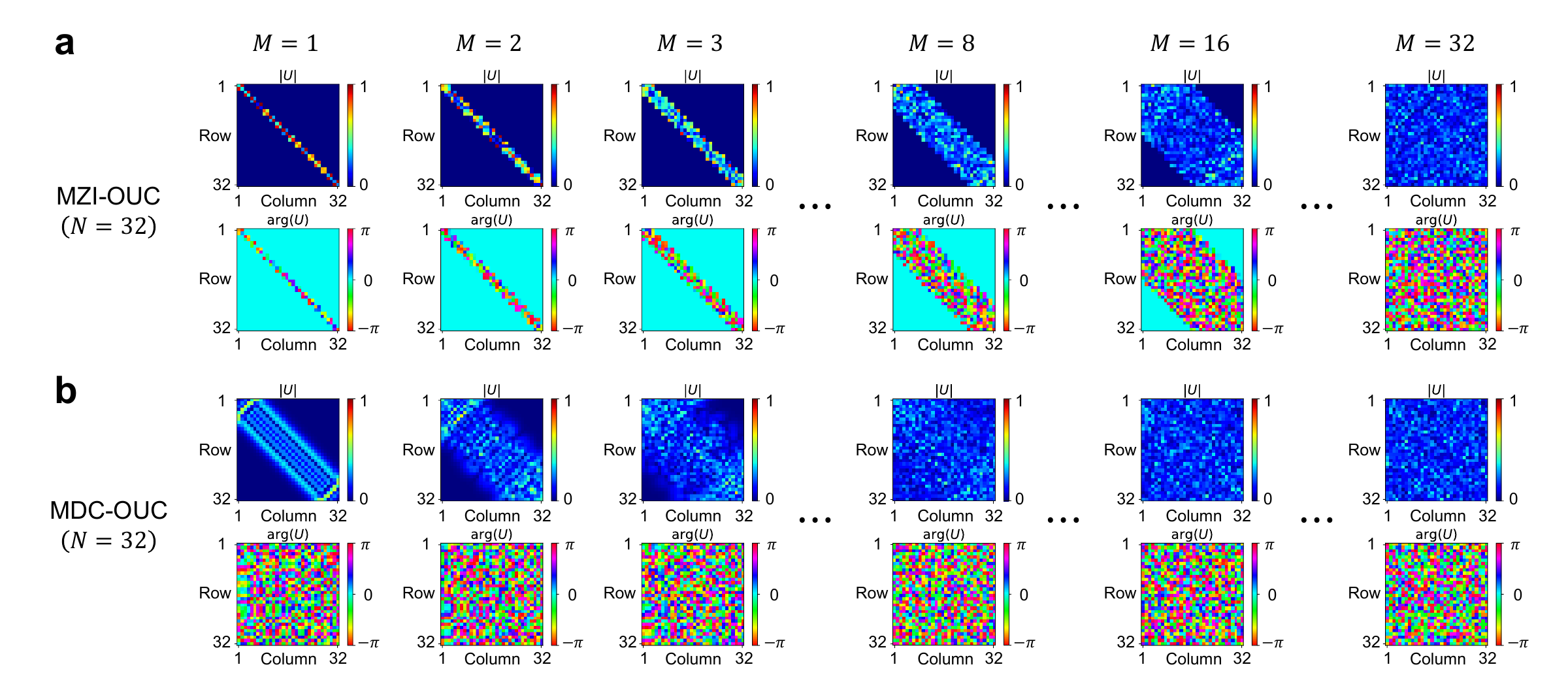}
\caption{
Random $32\times32$ unitary matrices generated by \textbf{a} MZI-OUC and \textbf{b} MDC-OUC  with various number of phase-shifting stages, $M$. MDC lengths are set to \SI{ 90 }{\micro m}.
}\label{Fig_S2}
\end{figure*}

Figure~\ref{Fig_S2} shows the absolute values and phases of all $32\times32$ components of an example matrix $U$ generated by MZI-OUC and MDC-OUC for $N=32$ and various values of $M$. 
We can confirm that the matrices generated by MDC-OUC tend to be denser than those generated by MZI-OUC even at small $M$, e.g. $M \sim 3$. 
This feature reflects the different light coupling natures of MDC and MZI; an MDC provides strong coupling among more than ten waveguides in a single stage (see Fig.~\ref{Fig_S1}), whereas an MZI only offers local coupling between adjacent ports. 

\begin{figure*}
\centering
\includegraphics[width=\textwidth]{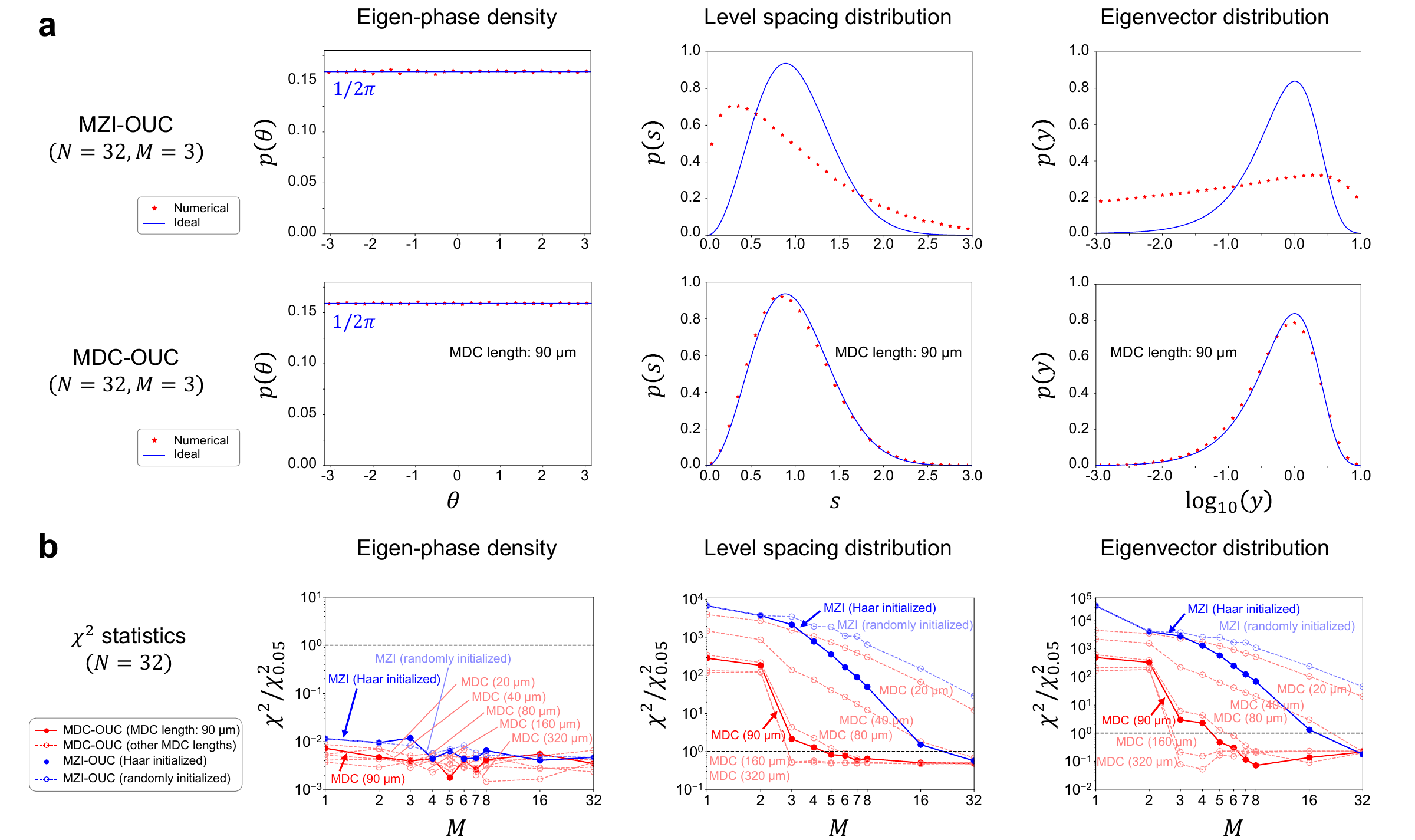}
\caption{
\textbf{a} Probability densities of three metrics representing the statistical distributions of eigenvalues and eigenvectors for random matrices generated by MZI-OUC and MDC-OUC with $N=32$ and $M=3$. The blue lines show the ideal Haar random cases, while the red dots represent the distributions of numerically generated unitary matrices. 
\textbf{b} Randomness of three metrics for MZI-OUC and MDC-OUC with $N=32$ and various $M$. The filled markers of $M=3$ corresponds to the plots shown in \textbf{a}. The highlighted lines of level spacing distribution corresponds to those shown in Fig.~\ref{Fig2}e in the main text.
}\label{Fig_S3}
\end{figure*}

\subsection{Haar randomness analysis}
\label{sec:suppl_1C}

To assess the randomness of the matrices generated by $N \times N$ OUCs, we analyze the probability distributions of eigenvalues and eigenvectors based on the random matrix theory \cite{mehta2004random,mezzadri2006generate,zyczkowski1994random}. Here, we use three metrics for evaluation: (1) the probability density of phases of $N$ eigenvalues (eigen-phases), (2) the level spacing distribution, and (3) the distribution of the squared absolute values of eigenvector components. The eigenvalues of a unitary matrix always have a form of $e^{j\theta}$, and the eigen-phase $\theta$ is theoretically expected to be uniformly distributed in the range of $[0,2\pi)$ \cite{mezzadri2006generate}. In other words, the probability density should satisfy
\begin{equation}
    p(\theta) = \frac{ 1 }{ 2 \pi } ~ ~.
    \label{eq:suppl_Haar_random_metric_1}
\end{equation}
We define the eigen-phases $\theta_i ~ (i=1,2,\dots,N)$ of an $N \times N$ unitary matrix in an ascending order, i.e. $\theta_1 \leq \theta_2 \leq \cdots \leq \theta_N$. The level spacing $s_i$ between the eigen-phases is then defined by 
\begin{equation}
    s_i \equiv \frac{N}{2\pi} ( \theta_{i+1} - \theta_i ) ~ ~.
\end{equation}
For Haar random unitary matrices, $s_i$ should distribute with a probability density function $p(s)$ in a form known as Wigner surmise \cite{mezzadri2006generate,zyczkowski1994random}:
\begin{equation}
    p(s) = \frac{ 32 s^2 }{ \pi^2 } \exp{
    \left( - \frac{ 4 s^2 }{ \pi } \right)
    }~.
    \label{eq:suppl_Haar_random_metric_2}
\end{equation}
Eigenvectors of $N \times N$ unitary matrix are defined as $v_i=\begin{pmatrix} v_{i1},v_{i2},\dots,v_{iN} \end{pmatrix}^\top (i=1,2,\dots,N)$. Then, the squared absolute values (defined as $y_{ij} \equiv \left| v_{ij} \right|^2)$ of all the eigenvector components should follow a certain distribution described as \cite{zyczkowski1994random}
\begin{equation}
    p(y) =
    c~
    \frac{ (\nu/2)^{\frac{\nu}{2}} }{ \Gamma(\nu/2) }
    \left( \frac{ y }{ \langle y \rangle } \right)^{ \frac{\nu}{2} - 1 }
    \exp{
    \left( - \frac{ \nu }{ 2 } \cdot \frac{ y }{ \langle y \rangle } \right)
    }~,
    \label{eq:suppl_Haar_random_metric_3}
\end{equation}
where $\nu = 4$ and $ \langle y \rangle $ represents the mean value of $y$. The scaling factor $c$ is set to 1.17 in our analysis.

Figure~\ref{Fig_S3}a shows the results for 20,000 random unitary matrices generated by MZI-OUC and MDC-OUC with $N = 32$ and $M = 3$. The blue lines show the probability densities of ideal Haar random cases calculated by Eqs.~(\ref{eq:suppl_Haar_random_metric_1})-(\ref{eq:suppl_Haar_random_metric_3}). The red dots represent the numerical distributions of each metric for the generated random unitary matrices. We can confirm that the unitary matrices generated by MZI-OUC are far from Haar random due to insufficient $M$. In contrast, those for MDC-OUC almost follow the Haar prediction, providing conclusive evidence of sufficient randomness realized by MDC-OUC with only $M = 3$, which is substantially smaller than $N = 32$.

To enable quantitative comparison, we further use chi-squared statistics to evaluate the conformity (goodness of fit) of the calculated distributions of MZI-OUC and MDC-OUC to that of Haar random unitaries. Given a series of observed data $f_i$ and expected data $p_i ~(i=1,\dots,K)$, the chi-squared statistics $ \chi^2 $ is defined as
\begin{equation}
    \chi^2 \equiv \sum_{i=1}^K \frac{ (f_i - p_i)^2 }{ p_i } ~.
\end{equation}
The smaller the $\chi^2$ value, the closer the calculated distribution is to the ideal case of Haar random unitaries. Furthermore, since the significance level of 5\% is $\chi^2_{0.05}(K-1)$ based on Pearson's chi-squared test, it is regarded to be far from the ideal distribution of Haar random unitaries if $\chi^2 > \chi^2_{0.05}(K-1)$. 
As a side note, the numerically calculated OUC distributions are discretized into $K=30$ uniformly spaced bins. Accordingly, the critical value at the 5\% significance level is $\chi^2_{0.05}(K-1) \simeq 42.56$. We normalized the chi-squared statistics to be $\chi^2/\chi^2_{0.05}$, so that when the value exceeds 1, the null hypothesis: ``each numerical distribution follows the ideal distribution of Haar random unitaries," will be rejected. 

From Fig.~\ref{Fig_S3}b, we can see that MZI-OUCs with $M\leq16$ are far from the ideal Haar random case, while MDC-OUC with $M=3$ can already generate matrices close to Haar random distribution ($\chi^2/\chi^2_{0.05} < 1$) when each MDC is long enough ($\gtrsim$ \SI{90}{\micro m}) to provide sufficient coupling. The $\chi^2/\chi^2_{0.05}$ values of MDC-OUC remain below 1 when $M\gtrsim3$, suggesting that MDC-OUC has high capability in generating sufficiently random unitary matrices that are close to Haar random. In addition, when the MDC is shorter (e.g. \SI{20}{\micro m}), chi-squared values approach to the case of MZI-OUC with uniformly random phase shifts, in which the directional coupler length is set to \SI{18.8}{\micro m} to satisfy $50:50$ power splitting condition, which is reasonable. This trend highlights the importance of strong nonlocal coupling required to achieve high randomness. 

\begin{figure*}[h!]
\centering
\includegraphics[width=\textwidth]{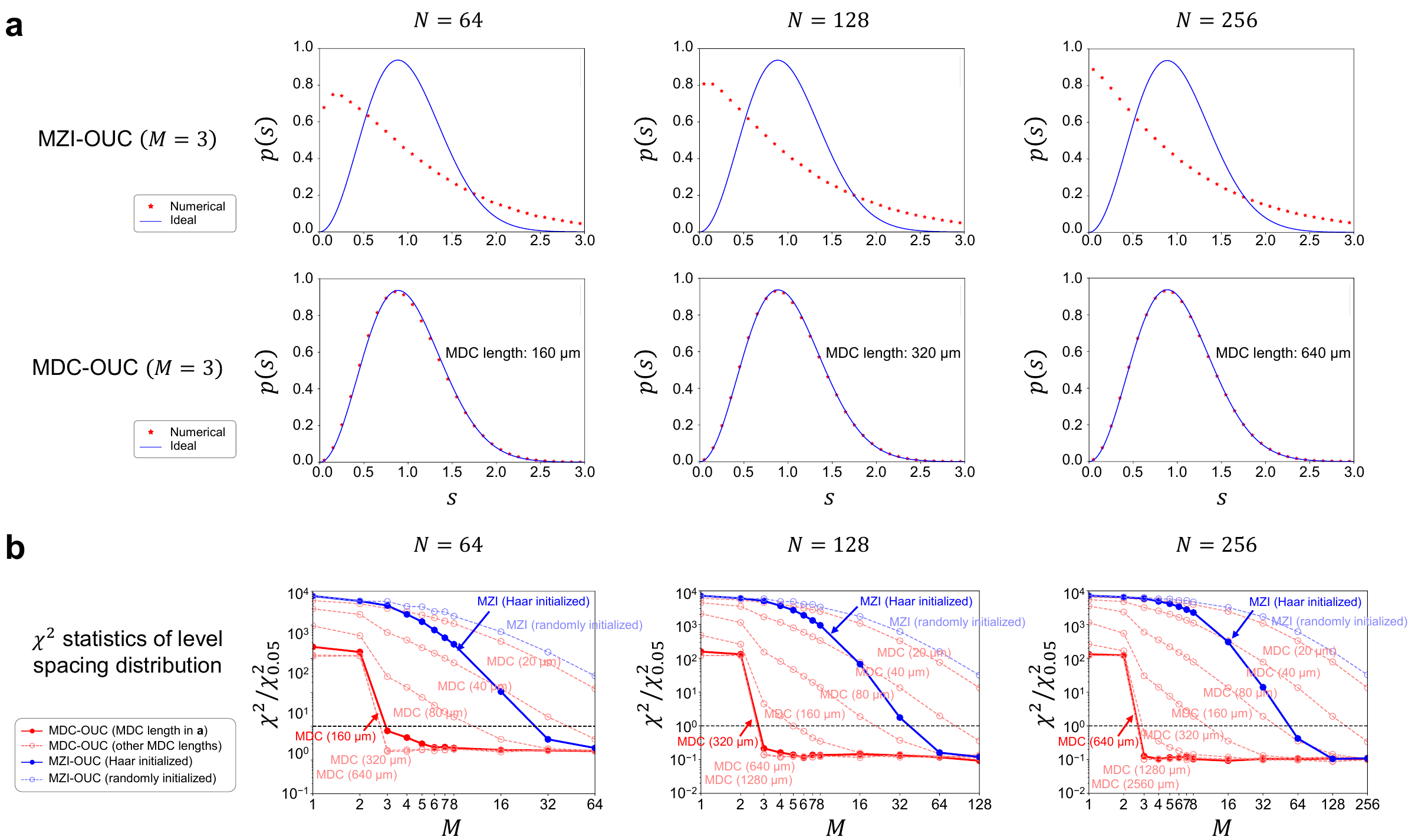}
\caption{
\textbf{a} Level spacing distributions of eigen-phases for the random matrices generated by MZI-OUC and MDC-OUC with $M=3$. The lengths of MDCs are set to $160$, $320$, and \SI{640}{\micro m} for $N=64$, $128$, and $256$, respectively.
\textbf{b} Randomness of the level spacing for MZI-OUCs and MDC-OUCs with $N=64,128,256$ and various $M$. The filled markers of $M=3$ corresponds to the plots shown in \textbf{a}.
}\label{Fig_S4}
\end{figure*}

Finally, we assess the scalability of MZI-OUCs and MDC-OUCs to larger $N$. Figure~\ref{Fig_S4}a shows the results of $N = 64$, $128$, and $256$, while $M$ is fixed to 3. We can confirm that superior randomness is obtained by MDC-OUC for all $N$. Figure~\ref{Fig_S4}b shows the dependence of $\chi^2$ value as a function of $M$ for each $N$. Surprisingly, steep drops in $\chi^2$ below $\chi^2_{0.05}$ are consistently observed at $M = 3$ for an arbitrarily large $N$, up to 256 that we could test using our computational resource. These results suggest that the required number of phase-shifting stages does not seem to scale with $N$, but is rather independent of $N$. This implies that sufficiently random unitary matrix with larger matrix sizes can be realized by using an MDC-OUC with only $ \sim 3N $ phase shifters in total, substantially improving the scalability compared to conventional MZI-OUC, which requires $N^2$ phase shifters.

\subsection{ONN simulation of MNIST and Fashion MNIST classification tasks}
\label{sec:suppl_1D}

\begin{figure*}[h]
\centering
\includegraphics[width=\textwidth]{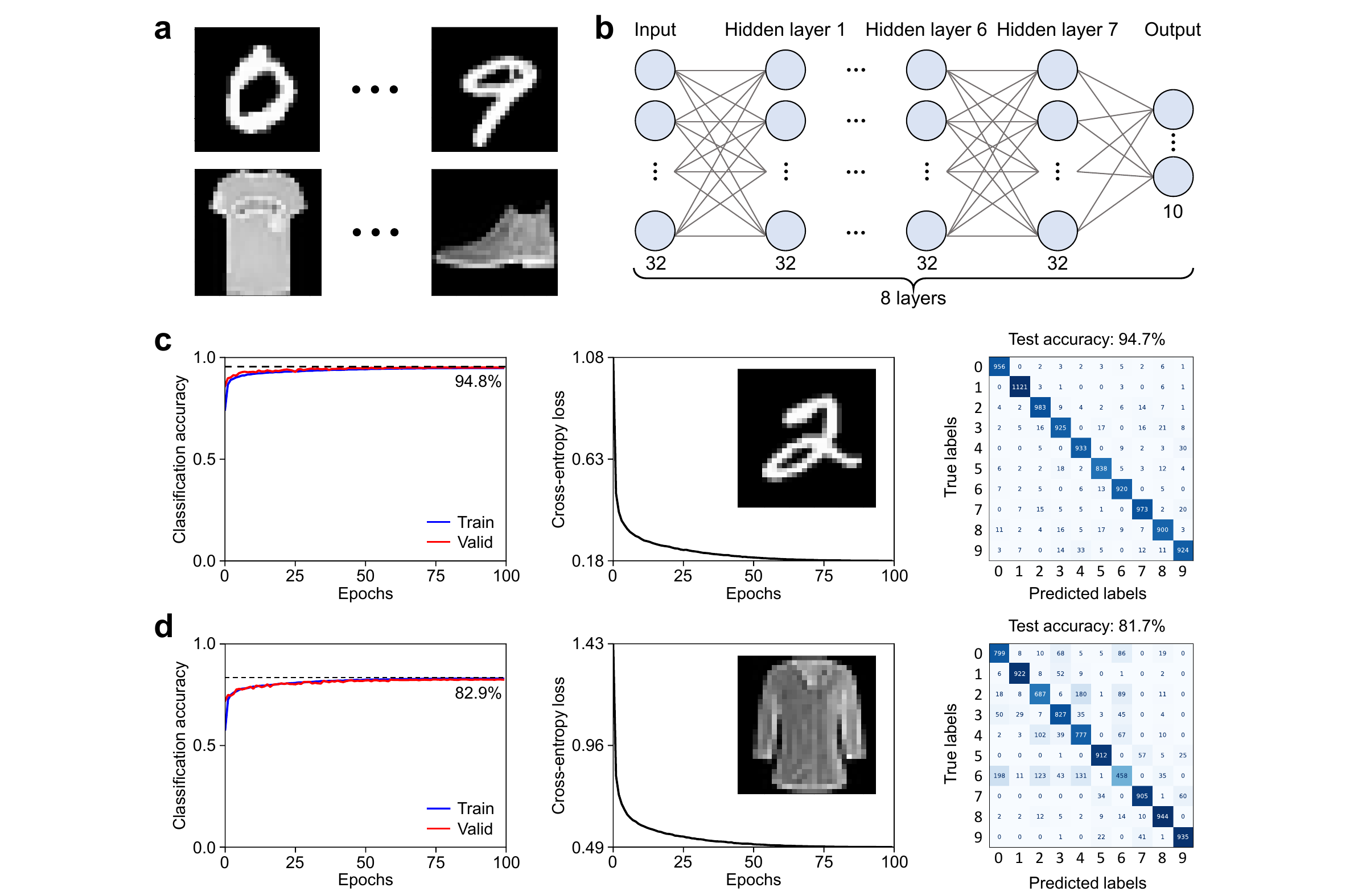}
\caption{
\textbf{a} Examples of images included in MNIST and Fashion MNIST dataset.
\textbf{b} Schematic of simulated ONN. We assume using our fabricated 32-input chip to realize each fully connected layer.
\textbf{c} Simulated results of classifying MNIST images. The training, validation accuracies and the loss against epochs, along with the confusion matrix of the testing data, are shown.
\textbf{d} Simulated results of classifying Fashion MNIST images.
}\label{Fig_S5}
\end{figure*}

The ONN performances of classifying MNIST \cite{lecun2002gradient} and Fashion MNIST \cite{xiao2017fashion} images are simulated. The sample images included in each dataset are shown in Fig.~\ref{Fig_S5}a. Both datasets are divided into 50,000 training data, 10,000 validation data, and 10,000 testing data. Figure~\ref{Fig_S5}b shows the architecture of simulated ONN, in which the number of layers is set to 8 for better classification accuracy, in contrast to the two-layer configuration used in our experiments. 
Assuming the use of our fabricated chip (Fig.~\ref{Fig3} in the main text), each fully connected layer consists of two 32-input 3-stage MDC-OUC with an MDC length of \SI{90}{\micro m} and an MZM array that applies singular values. All fully connected layers are followed by ReLU as the activation function, except for the output of the last layer, which is fed to Softmax for calculating the cross-entropy loss. We modeled the fabricated chip using Eq.~(\ref{eq:U_MDC}) and trained the ONN by backpropagation algorithm.

The simulated results are shown in Fig.~\ref{Fig_S5}c. The ONN achieves a test accuracy of 94.7\%, which is close to the training accuracy of 94.8\%. Figure~\ref{Fig_S5}d presents the simulated results of Fashion MNIST classification. Similarly, the ONN achieves a test accuracy of 81.7\%, showing a good match with the training accuracy of 82.9\%.

\begin{table*}[ht]
\caption{MNIST classification accuracy by two-layer ONN with 3-stage OUCs.}\label{tab:suppl_MNIST_scalability}
\begin{tabular*}{\linewidth}{ @{\extracolsep{\fill}} llllll }
\toprule
OUC type ($M=3$) & $N=32$ & $N=64$ & $N=128$  & $N=256$ & $N=512$ \\
\midrule
MDC-OUC  &  93.3\%  &  95.7\%  &  97.0\%  &  97.8\%  &  98.6\%  \\
MZI-OUC  &  83.3\%  &  76.6\%  &  64.4\%  &  48.4\%  &  35.5\%  \\
\botrule
\end{tabular*}
\end{table*}

\begin{table*}[ht]
\caption{Fashion MNIST classification accuracy by four-layer ONN with 3-stage OUCs.}\label{tab:suppl_Fashion_MNIST_scalability}
\begin{tabular*}{\linewidth}{ @{\extracolsep{\fill}} llllll }
\toprule
OUC type ($M=3$) & $N=32$ & $N=64$ & $N=128$ & $N=256$ & $N=512$ \\
\midrule
MDC-OUC  &  82.8\%  &  86.8\%  &  88.2\%  &  89.1\%  &  89.6\%  \\
MZI-OUC  &  77.1\%  &  76.6\%  &  65.6\%  &  57.1\%  &  42.7\% \\
\botrule
\end{tabular*}
\end{table*}

The scalability of ONN is also compared for the cases of using MDC-OUCs and MZI-OUCs with $M=3$ and increasing $N$. Here, the length of MDC is set to \SI{10}{\micro m} $\times N$ to ensure sufficient all-to-all coupling for all cases of $N$. The number of layer is set to 2 for consistency with our experimental results and due to the memory limitation of our computing resource for $N=512$.
The simulated accuracies for MNIST classification task using two-layer ONN of various $N$ are summarized in Table~\ref{tab:suppl_MNIST_scalability}. The test accuracy using MDC-OUCs improves steadily from 93.3\% to 98.6\% as $N$ increases from 32 to 512. 
In contrast, it decreases monotonically from 83.3\% to 35.5\% with increasing $N$ for the case of MZI-OUC. 
Similarly, Table~\ref{tab:suppl_Fashion_MNIST_scalability} shows the performance of Fashion MNIST classification task using four-layer ONN.
Once again, the accuracy improves continuously for the case of MDC-OUC, whereas it decreases monotonically for MZI-OUC as $N$ increases.
This difference is attributed to the fact that sufficient nonlocal coupling is provided by MDC-OUC, whereas the port-to-port connectivity drastically degrades for MZI-OUC at large $N$ and small $M$. 
Owing to the excellent randomness and expressivity of MDC-OUC, we can achieve higher ONN performance by increasing $N$ while fixing $M$ to 3, implying that the required number of phase shifters scales at $O(N)$ instead of $O(N^2)$.

\vspace{1.0ex}% adjust spacing
\section{Experiments}
\label{sec:suppl_2}
\vspace{-5.0ex}% adjust spacing
\subsection{Characterization of a single phase shifter and photodetector}
\label{sec:suppl_2A}

\begin{figure*}[ht]
\centering
\includegraphics[width=\textwidth]{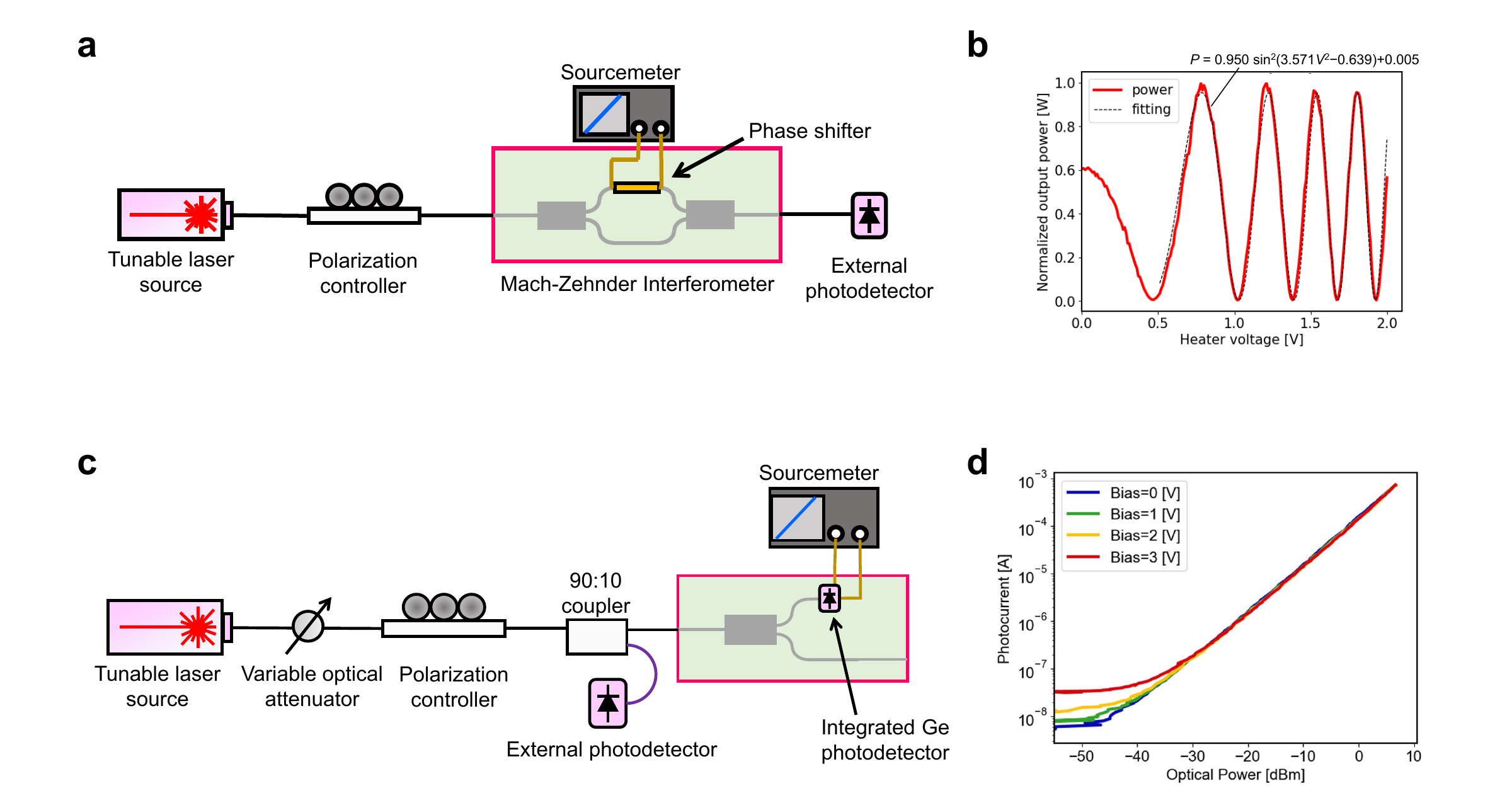}
\caption{
Characterization of a single phase shifter and PD of test patterns. 
\textbf{a} Schematic of the setup of measuring a test phase shifter. 
\textbf{b} Measured characteristic of the test phase shifter. 
\textbf{c} Schematic of the setup of measuring a test PD. 
\textbf{d} Measured characteristic of the test PD. 
}\label{Fig_S9}
\end{figure*}

We characterized the test phase shifter and PD, which were fabricated on the same chip. The experimental setups and measured results are shown in Fig.~\ref{Fig_S9}. 
The phase shifter was characterized by applying voltage to the phase shifter attached on a test Mach-Zehnder interferometer via a sourcemeter (Keithley 6487) as shown in Fig.~\ref{Fig_S9}a. The input light is fiber-coupled to the chip, and the output optical power is detected by an external photodetector. 
From the measured electrical resistance (see Fig.~\ref{Fig4}a in the main text), $\pi$ phase shift is obtained at 1.23 mW. 
Next we measured the photocurrent from the PD integrated on the chip, by applying reverse bias using the sourcemeter. We used a 90:10 coupler to monitor the input optical power fiber-coupled to the chip. For different bias voltages of PD ranging between \SI{ 0 }{ V } and \SI{ 3 }{ V }, the responsivity is derived to be \SI{ 0.75 }{ A/W }.

\subsection{Experimental setup}
\label{sec:suppl_2B}

\begin{figure*}[ht]
\centering
\includegraphics[width=\textwidth]{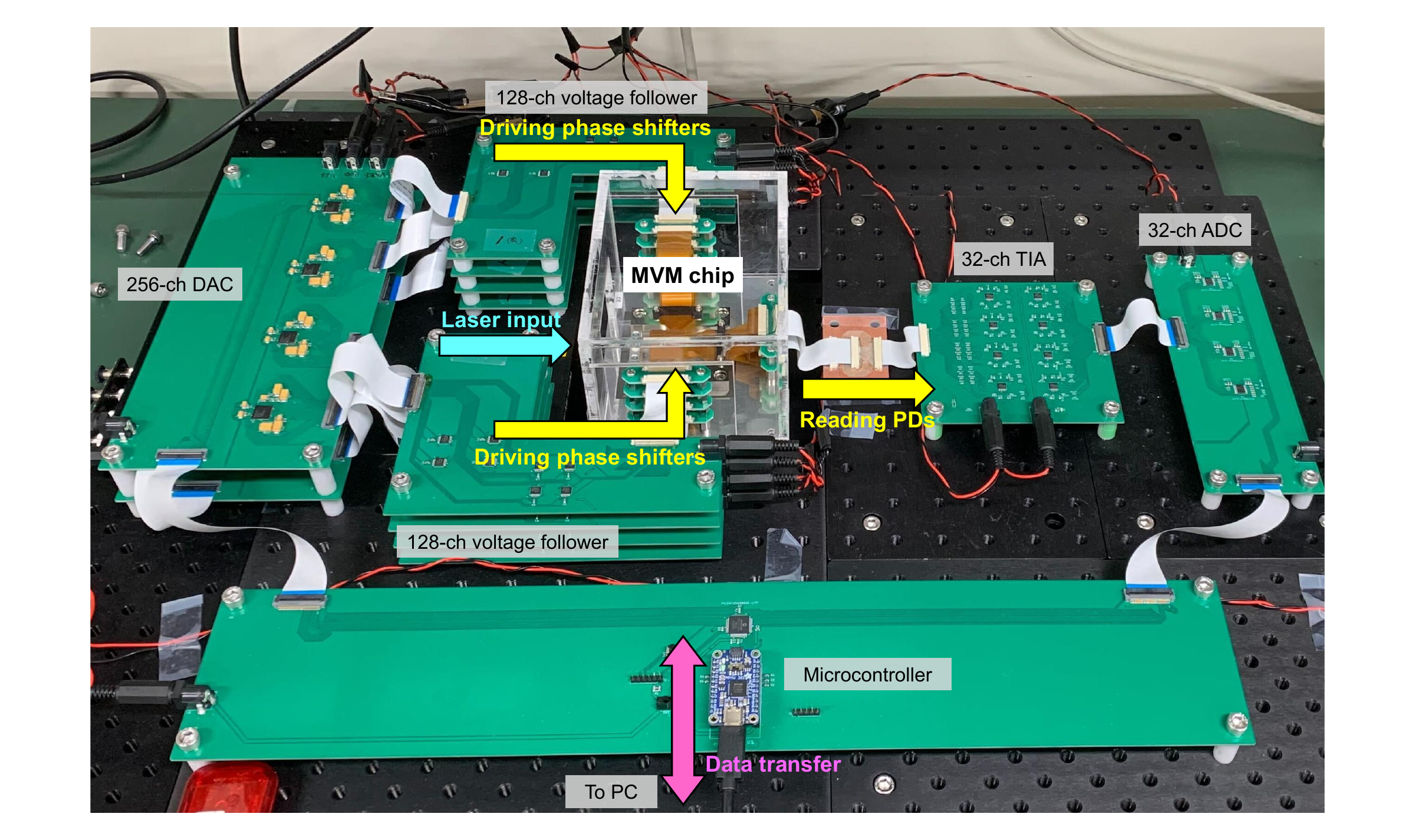}
\caption{
Photograph of the experimental setup. Custom PCBs are used for driving all 256 phase shifters and reading 32 photodetectors integrated on the MVM chip. A microcontroller serves as an I/O unit for transferring data from the PC to DACs and from ADCs to the PC.
}\label{Fig_S6}
\end{figure*}

The fabricated MVM chip contained 256 phase shifters and 32 PDs, requiring electrical control of 288 active components and stable fiber coupling, which necessitated the development of custom packaging for stable laboratory experiments (see Figs.~\ref{Fig3}a and \ref{Fig3}f in the main text). 
All on-chip components were connected to an external electric controller through 324 bonded wires and 10 flexible cables. 
We developed custom printed circuit boards (PCBs) containing 256-channel D/A converters (DAC, AD5370, 40-channel and 16-bit precision), each connected to a voltage follower circuit to supply sufficient current for driving all 256 phase shifters as shown in Fig.~\ref{Fig_S6}. 
The light from a tunable laser source (Keysight 81980A) was input to the MVM chip. The output optical signals were detected by 32 on-chip PDs, converted to photocurrents, amplified by 32-channel transimpedance amplifiers (TIAs), and processed by A/D converters (ADCs, MCP3564R, 8-channel and 24-bit precision). 
The DACs and ADCs were all controlled by a microcontroller (PIC24FJ256GB606) through SPI communication to synchronize input vector generation and signal readout. The microcontroller was connected to a serial port module (FT232H), supporting Python-based programming via USB connection to a computer.

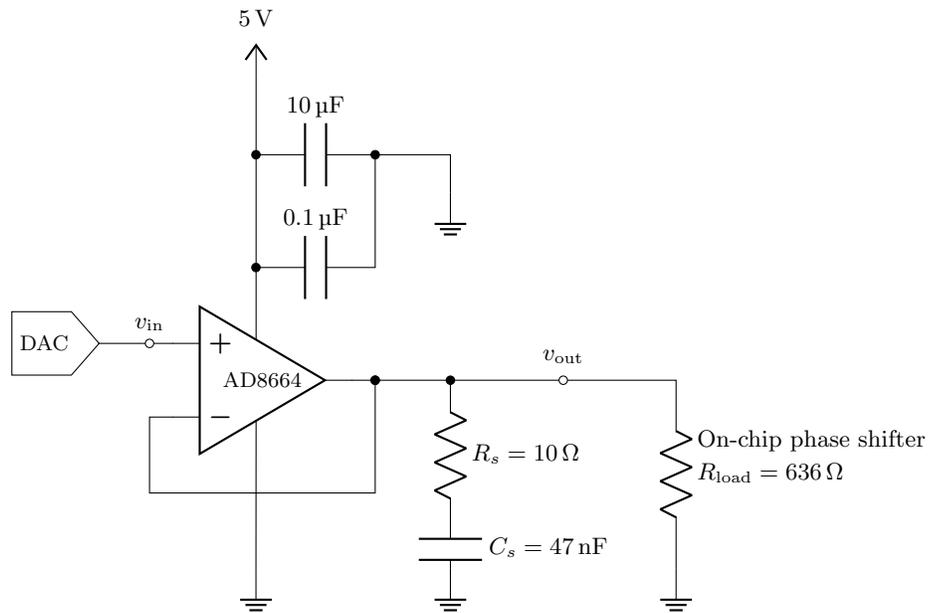
\begin{figure}[ht]
	\centering
	\scalebox{1.0}{ % graphicx
		\begin{circuitikz}
			\draw[white]
			(-3.5, 0.99) to[generic,color=white,name=DAC] (-1, 0.99);
			\myDAC{DAC}
			;
			\draw
			(-1.68, 0.99) to[short] (-1, 0.99) node[]{}
			;
			\draw
			(4.5, 0.5)
			to[short] (6, 0.5)
			to[R, label=\mbox{$ R_{ \mathrm{load} } = \SI{ 636 }{ \Omega } $}] (6, -2) node[ground]{}
			;
            \node[align=left] at (7.8, -0.3) {On-chip phase shifter}
            ;
			\draw
			(0.5, 0.5) node[op amp, noinv input up, label={[label distance=-6pt]above:\footnotesize{AD8664}}] (opamp) {}
			(opamp.+) to[short] (-1, 0.99) node[ocirc, label={above:$v_{\mathrm{in}}$}]{}
			(opamp.-) to[short] (-1, 0.01) to[short] (-1, -1) to[short] (2, -1) to[short, -*] (2, 0.5)
			(opamp.out) to[short, -*] (2, 0.5) to[short, -o] (4.5, 0.5) node[label={above:$v_{\mathrm{out}}$}]{}
			(opamp.up) --++(0, 3.5) node[vcc]{ \SI{ 5 }{ V } }
			(opamp.down) to[short] (0.415, -2) node[ground]{}
			;
			\draw
			(3, 0.5)
			to[R, label=\mbox{$ R_s = \SI{ 10 }{ \Omega } $}, *-] (3, -1.5)
			to[C, label=\mbox{$ C_s = \SI{ 47 }{ nF } $}] (3, -2) node[ground]{}
			;
			\draw
			(0.415, 2.0) to[C, label=\SI{ 0.1 }{ \micro F }, *-] (2, 2.0) to[short, -*] (2, 3.5)
			(0.415, 3.5) to[C, label=\SI{ 10  }{ \micro F }, *-] (2, 3.5)
			to[short] (3, 3.5) to[short] (3, 3.0) node[ground]{}
			;
		\end{circuitikz}
	}
	\caption{Schematic of a voltage follower with bypass capacitors and snubber circuit.}
	\label{cir:voltage_follower}
\end{figure}

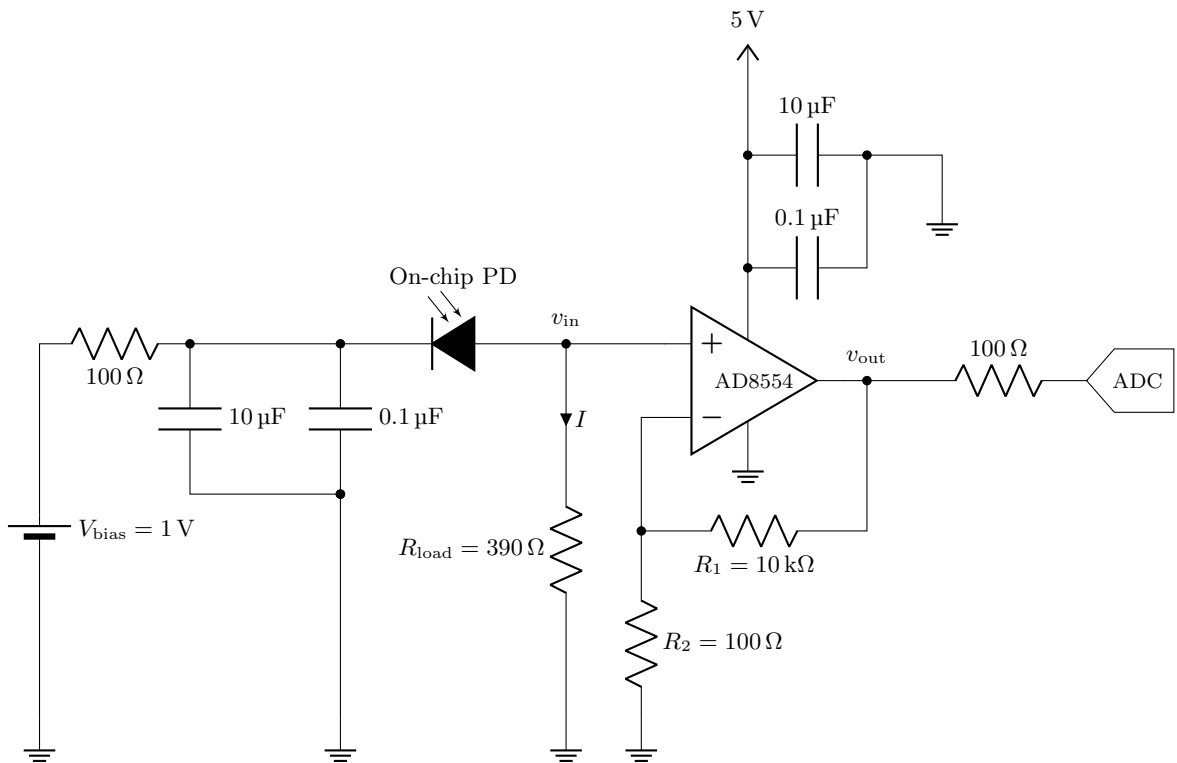
\begin{figure}[ht]
	\centering
	\scalebox{1.0}{
		\begin{circuitikz}
			\draw
			(0.5, 0.5) node[op amp, noinv input up, label={[label distance=-6pt]above:\footnotesize{AD8554}}] (opamp) {}
			(opamp.+) to[short, -*] (-2, 0.99) node[label={above:$v_{\mathrm{in}}$}]{} to[short, i=$ I $] (-2, -1) to[R, l_=\mbox{$ R_{ \mathrm{load} } = \SI{ 390 }{ \Omega } $}] (-2, -2.5) to[short] (-2, -4) node[ground]{}
			(opamp.-) to[short] (-1, 0.01) to[short, -*] (-1, -1.5)
			(opamp.out) to[short, -*] (2, 0.5) node[label={above:$v_{\mathrm{out}}$}]{} to[short] (3, 0.5) to[R, label=\SI{ 100 }{ \Omega }] (4.5, 0.5)
			(opamp.up) --++(0, 3.5) node[vcc]{ \SI{ 5 }{ V } }
			(opamp.down) --++(0, -0.25) node[ground]{}
			;
			\draw
			(0.415, 2.0) to[C, label=\SI{ 0.1 }{ \micro F }, *-] (2, 2.0) to[short, -*] (2, 3.5)
			(0.415, 3.5) to[C, label=\SI{ 10  }{ \micro F }, *-] (2, 3.5)
			to[short] (3, 3.5) to[short] (3, 3.0) node[ground]{}
			;
			\draw
			(2, 0.5) to[short] (2, -1.5)
			to[R, l=\mbox{$ R_1 = \SI{ 10 }{ k \Omega } $}] (-1, -1.5)
			to[short] (-1, -2)
			to[R, label=\mbox{$ R_2 = \SI{ 100 }{ \Omega } $}] (-1, -4) node[ground]{}
			;
			\draw
			(-2, 0.99)
			to[pD*, l_=On-chip PD, mirror] (-5, 0.99)
			to[short, *-*] (-7, 0.99)
			to[R, label=\SI{ 100 }{ \Omega } ] (-9, 0.99)
			to[battery2, label=\mbox{$ V_{ \mathrm{bias} } = \SI{ 1 }{ V } $}] (-9, -4) node[ground]{}
			;
			\draw
			(-5, 0.99) to[C, label=\SI{ 0.1 }{ \micro F }, -*] ++(0, -2) to[short] (-5, -4) node[ground]{}
			(-7, 0.99) to[C, label=\SI{ 10  }{ \micro F }] ++(0, -2) to[short, -*] ++(2, 0)
			;
			\draw[white]
			(4.5, 0.5) to[generic,color=white,name=ADC] (6.5, 0.5);
			\myADC{ADC}
			;
			\draw
			(4.5, 0.5) to[short] (4.93, 0.5) node[]{}
			;
		\end{circuitikz}
	}
	\caption{Complete schematic of readout circuit using TIA.}
	\label{cir:TIA}
\end{figure}

The custom voltage follower circuit was constructed by an operational amplifier (AD8664) as shown in Fig.~\ref{cir:voltage_follower}. The input voltage $v_{\mathrm{in}}$ from the DAC was forwarded to the output voltage $v_{\mathrm{out}}$. Here, $v_{\mathrm{out}}$ was stabilized by a snubber circuit constructed by $R_s$ and $C_s$, which formed a low-pass filter to suppress the oscillation of the amplifier. The output voltage $v_{\mathrm{out}}$ was applied to each phase shifter.
Figure~\ref{cir:TIA} shows the custom TIA circuit. 
The input voltage $v_{\mathrm{in}}$ generated at the load resistor $R_{\mathrm{load}}$ was sent to the operational amplifier (AD8554), which constructed a non-inverting amplifier circuit with a gain of $\frac{R_1+R_2}{R_2}$, thus $v_{\mathrm{out}} = 101 v_{\mathrm{in}}$. The output voltage $v_{\mathrm{out}}$ was detected by ADC and converted to a digital signal. 

\subsection{Characterization of active components}
\label{sec:suppl_2C}

\begin{figure*}[ht]
\centering
\includegraphics[width=\textwidth]{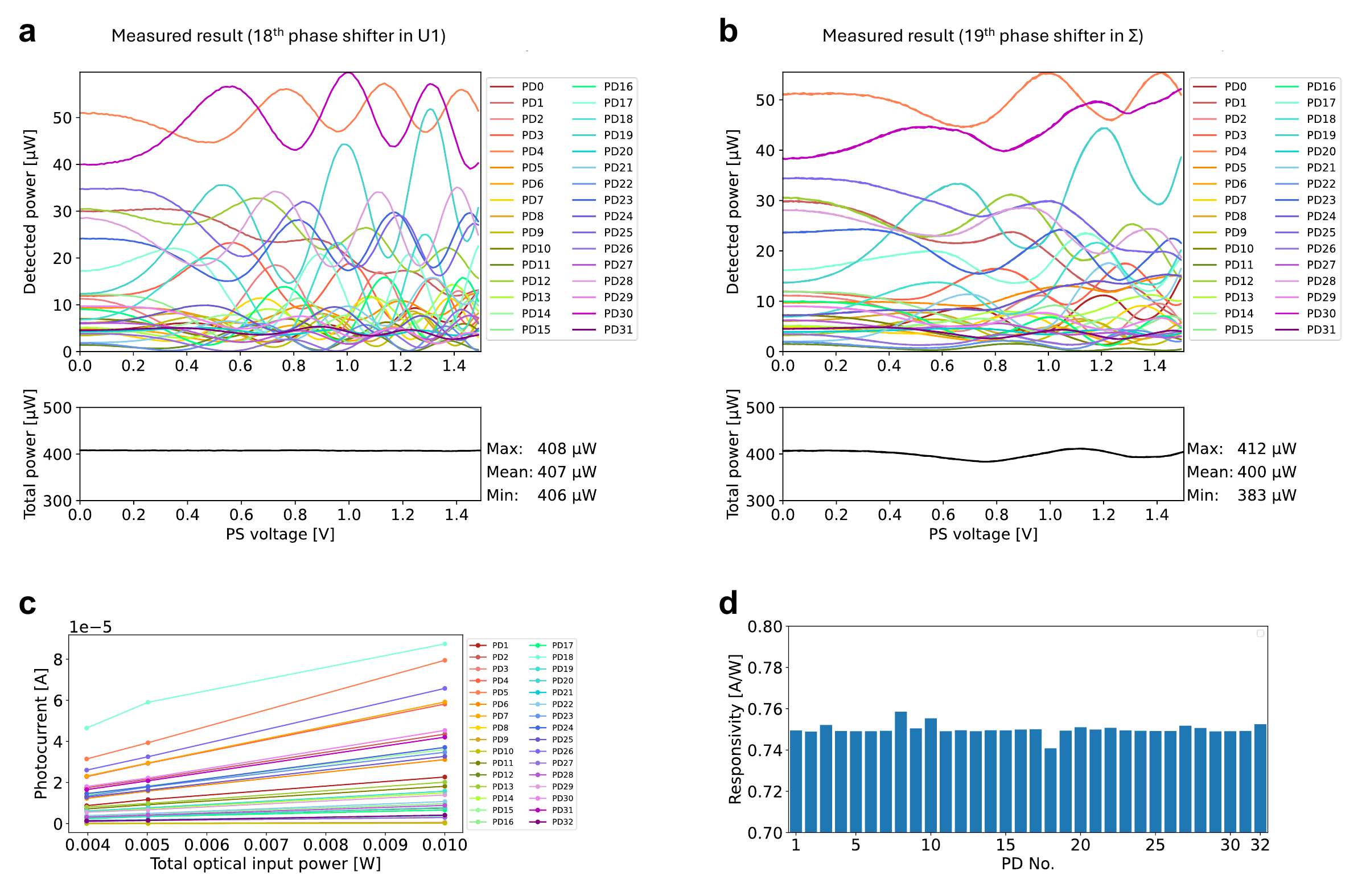}
\caption{
Characterization of phase shifters and PDs on the PIC. 
\textbf{a} Detected optical power when sweeping the applied voltage of the 18th phase shifter located in the first phase shifter array of MDC-OUC $U$. 
\textbf{b} Detected optical power when sweeping the applied voltage of the 19th phase shifter located in the MZM array $\Sigma$. 
\textbf{c} Measured photocurrent of PDs when changing the total optical laser power. 
\textbf{d} Responsivity of all PDs derived from \textbf{c}.
}\label{Fig_S10}
\end{figure*}

The phase shifters on the chip were characterized by sweeping the applied voltage to each phase shifter and observing the sinusoidal response of the photocurrents at 32 PDs. 
Figure~\ref{Fig_S10}a shows the measured results for an example case, where the 18-th phase shifter in the second OUC section is characterized.
We can confirm that the intensity distribution at 32 PDs varies periodically with the applied voltages.
The total optical power detected by all PDs (lower panel in Fig.~\ref{Fig_S10}a) is constant due to the lossless operation of phase shifters and unitary conversion of the following MDCs. 
In contrast, Fig.~\ref{Fig_S10}b shows the case of characterizing the 19-th phase shifter at the MZM array section.
Now, the total optical power also varies periodically corresponding to the intensity modulation by the MZM. From these results, we fitted the efficiency of each phase shifter by assuming a quadratic voltage-to-phase relationship as described in the Methods section of the main text.

The integrated PDs were also characterized. Figure~\ref{Fig_S10}c shows the measured photocurrent of all 32 PDs. The optical power variation among 32 PDs was calibrated using the measured photocurrent values at a fixed total input power of \SI{6}{dBm}. Then, from the slope of the measured points, we derived the responsivity of $n$-th PD via $R_n = \frac{dI_n}{dP_n}$.

\subsection{In-situ training using simulated annealing}
\label{sec:suppl_2D}

We employed simulated annealing for in situ training of ONN as shown in Fig.~\ref{Fig_S11}. We first initialized all weights to random values in the range $[0, 2\pi)$ and performed a feed-forward propagation of the input batch. We employed the same MVM chip to implement two-layer ONN configuration in this work, where the measured output from the first layer was used as the inputs to the second layer after applying the ReLU activation. Finally, a Softmax operation was applied to the measured signals to obtain the initial classification results, from which the cross-entropy loss was computed.
Then all weights were trained following the optimization loop as described in the Methods section of the main text. The pseudo code is shown in Algorithm~\ref{alg:SA}.

In the experiments, the dataset sizes were limited to a few hundred samples, primarily due to the latency of data transfer between electrical circuits of DACs and ADCs, which required several tens to hundreds of milliseconds to update the phase shifters and readout the photodetectors. Consequently, we conducted a proof-of-concept demonstration based on 2-class MNIST image classification rather than the full 10-class task. 
This bottleneck is not fundamental and can be improved by careful design of control circuitries and ultimately through the photonic-electrical co-packaging. By integrating high-speed modulators for the input vector generation, the clock rate can be further increased beyond the \si{GHz} regime, enabling high-throughput computation using substantially larger datasets.

% voltage follower circuit

% TIA circuit

\begin{figure}[ht]
\flushleft
\begin{minipage}{0.8\textwidth}
\refstepcounter{algorithmctr}
\caption*{\textbf{Algorithm 1.} Simulated annealing procedure.}
\label{alg:SA}
\vspace{-0.1cm}
\hrule
\vspace{0.1cm}
\begin{algorithmic}[1]
\State Initialize $\delta\phi \gets 0.1\pi$, $T \gets T_0$
\State Initialize all weights $\phi_{i,j,k}$
\Comment{$i$: port index, $j$: stage index, $k$: layer index}
\For{$n = 1$ to $\mathrm{batch\_size}$}
    \State Load input vector $x_n$
    \State Measure output $y_{i,n}$
    \State Compute 
    $q_{i,n} \gets 
    \dfrac{\exp(y_{i,n})}{\sum_{i'} \exp(y_{i',n})}$
\EndFor
\State Compute initial loss $L \gets -\sum_{n}\sum_{i} p_{i,n}\log q_{i,n}$
\Repeat
    \State Randomly select $(i,j,k)$
    \State $\phi_{\mathrm{prev}} \gets \phi_{i,j,k}$
    \State $\phi_{i,j,k} \gets \phi_{i,j,k} + \text{Uniform}(-\delta\phi,\delta\phi)$
    \For{$n = 1$ to $\mathrm{batch\_size}$}
        \State Load input vector $x_n$
        \State Measure output $y_{i,n}$
        \State Compute $q_{i,n}$
    \EndFor
    \State Compute new loss $L' \gets -\sum_{n}\sum_{i} p_{i,n}\log q_{i,n}$
    \If{$L' < L \;\lor\; \exp\!\left(-(L'-L)/T\right) > \text{Uniform}(0,1)$}
        \State $L \gets L'$
        \State $T \gets \alpha T$
        \Comment{$\alpha \in (0,1)$: cooling rate}
    \Else
        \State $\phi_{i,j,k} \gets \phi_{\mathrm{prev}}$
        \State $ \delta \phi \gets \delta \phi - \epsilon$
        \Comment{$\epsilon$: small decrement constant}
    \EndIf
\Until{$\delta\phi < 10^{-3}$ \textbf{and} $T < 10^{-10}$}
\end{algorithmic}
\vspace{0.1cm}
\hrule
\end{minipage}
\end{figure}

\begin{figure*}[ht]
\centering
\includegraphics[width=\textwidth]{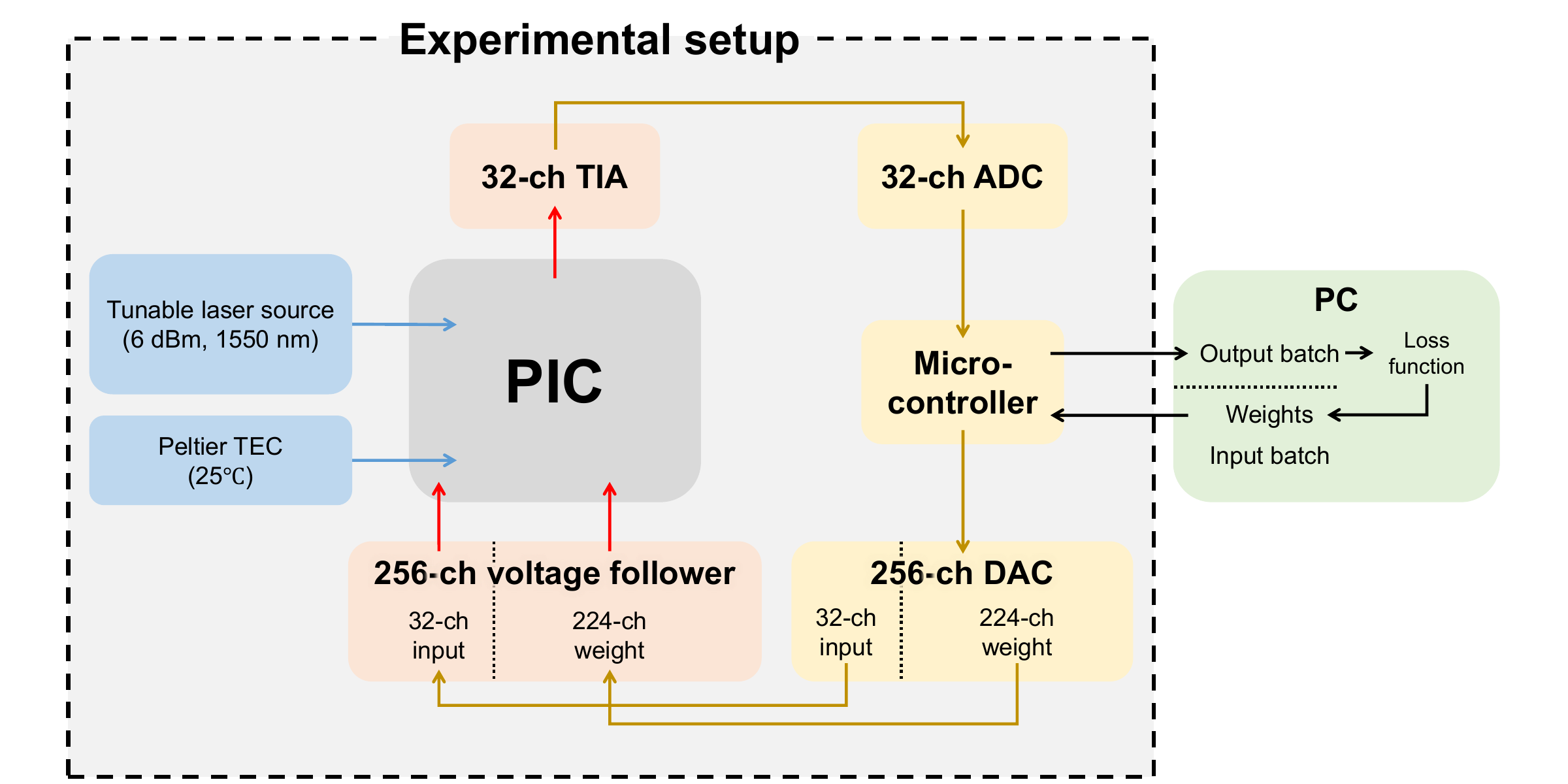}
\caption{
Schematic of the experimental setup and the in-situ training procedure of the ONN. In each iteration, a randomly selected weight is updated and loaded onto the PIC. The setup performs feed-forward propagation of the input batch, and the resulting output batch is measured and evaluated by the loss function. Then, we decide whether to accepted the updated weight by comparing the loss value with the previous value, after which the training proceeds to the next iteration.
}\label{Fig_S11}
\end{figure*}

\end{document}